\documentclass[a4paper, 11pt]{article}
\usepackage[pdftex, a4paper]{geometry}
\usepackage{fancyhdr}
\usepackage{amsfonts}
\usepackage{dsfont}
\usepackage{mathrsfs}
\usepackage{amssymb}
\usepackage{bbm}
\usepackage{epsfig}
\usepackage{paralist}
\usepackage{amsmath}
\usepackage{appendix,tabularx}
\usepackage{supertabular}
\usepackage[english]{babel}
\usepackage{cite}
\usepackage{color}
\usepackage{graphicx}
\usepackage{caption}
\usepackage{subcaption}
\DeclareMathAlphabet{\mathpzc}{OT1}{pzc}{m}{it}
\newtheorem{theorem}{Theorem}
\newtheorem{lemma}{Lemma}

\newtheorem{proposition}{Proposition}
\newtheorem{remark}{Remark}

\newtheorem{definition}{Definition}

\def\vec{\mathrm{vec}}
\def\sfN{\mathsf{N}}
\def\sfM{\mathsf{M}}

\def\diag{\mathrm{diag}}
\def\Tr{\mathrm{Tr}}
\def\bb{\mathbb}
\def\ker{\mathrm{Ker}}

\begin{document}

\title{\bf Sampled-Data Consensus over Random Networks}
\date{}

\author{Junfeng Wu, Ziyang Meng, Tao Yang,
Guodong Shi, and Karl H. Johansson}
\maketitle

\begin{abstract}
This paper considers the consensus problem for a network of nodes with random interactions and sampled-data control actions.
We first show that consensus in expectation, in mean square, and almost surely
are equivalent for a general random network model when the inter-sampling interval and network
size satisfy a simple relation.
The three types of consensus are shown to be simultaneously achieved
over an independent or a Markovian random network defined
on an underlying graph with a directed spanning tree.
For both independent and
Markovian random network models, necessary and sufficient conditions for mean-square consensus are derived in terms of the spectral radius of the corresponding state transition matrix. These conditions are then interpreted as the existence of critical value on the inter-sampling interval, below which global mean-square consensus is achieved and above which the system diverges in mean-square sense for some initial states.
Finally, we establish an upper bound on the
inter-sampling interval below which almost
sure consensus is reached,
and a lower bound on the inter-sampling
interval above which almost sure divergence
is reached. Some numerical simulations are given to validate the theoretical results and some discussions on the critical value of the inter-sampling intervals for the mean-square consensus are provided.
\end{abstract}


%

\section{Introduction}
In the traditional consensus algorithm, each node exchanges information with a few neighbors, typically given by their relative states, and then updates its own state according to a weighted average. It turns out that with suitable (and rather general) connectivity conditions imposed on the communication graph, all nodes asymptotically reach an agreement in which the nodes' initial values are encoded \cite{JadbabaieLinMorse03,SaberMurray04}.
Various consensus algorithms have been proposed in the literature.
The most
common continuous-time consensus algorithm is given by an ordinary differential equation in terms
of the relative states of each agent with respect to its neighboring agents~\cite{SaberMurray04,LinZhiyun_SIAM07}.
The agent state is driven towards the states of its neighbors, so eventually the algorithm ensures that
the whole network reaches an agreement provided that the network is jointly connected.
In \cite{Boyd_SCL04,RenBeard05_TAC}, the authors developed discrete-time consensus algorithms. In such an algorithm,
each agent updates its states as a convex combination of the state of itself and that of its neighboring agents.
Due to the fact that most algorithms are implemented by a digital device and
that the communication channels are unreliable and often subject to limited transmission capacity,
sampled-data consensus algorithms have also been proposed~\cite{CaoYongcan_IJRNC10,gao2011sampled,zhang2010consensus,Feng-TAC12}. In a sampled-data setting, the agent dynamics are continuous and the control input is piecewise continuous. The closed-loop system is transformed into
 discrete-time dynamics and conditions on uniform or nonuniform sample periods are critical to ensure consensus.

Consensus over random networks has drawn much attention since communication networks are naturally random.
In \cite{Kar-TSP08,Kar-TSP09}, the authors studied distributed average consensus
in sensor networks with quantized data and independent, identically distributed (i.i.d.) symmetric random topologies.
The authors of \cite{pereira-TSP2010} evaluated the mean-square convergence of consensus algorithms with random asymmetric topologies.
Mean-square performance for consensus algorithms over i.i.d. random graphs was studied in \cite{fagnani2008randomized}, and the impact of random packet drops was investigated in \cite{Fagnani-SIAM}.
Recently, the i.i.d. assumption was relaxed in \cite{ali-TAC10,Baras-SIAM13} to the case where the communication graph is modeled by a finite-state Markov chain. Probabilistic consensus has also been investigated in the literature. It was shown in
\cite{chen-TAC2011} that for a random network generated by i.i.d. stochastic matrices, almost sure, in probability, and ${L}^p$ ($p\geq 1$) consensus are equivalent.
In \cite{mesbahi-TAC05}, the authors showed that almost sure convergence is reached for i.i.d. random graphs, and Erd\H{o}s-R\'enyi random graphs.
The analysis was later extended to directed graphs and more general random graph processes
\cite{Wu_TAC06,Stilwell-TAC07}. In \cite{ali-TAC08}, the authors showed that asymptotic almost sure consensus over i.i.d. random networks is reached if and only if the graph contains a directed spanning tree in expectation.
Divergence in random consensus networks has also been considered, as representing asymptotic disagreement in social networks. Almost sure
divergence of consensus algorithms was considered
in~\cite{acemoglu2013opinion,shi2013agreement}.

In this paper, we consider sampled-data consensus problems over random networks.
We analyze the convergence of the consensus algorithm with
a sampled-data controller under two random network models. In the first model,
each node independently samples its neighbors in a random manner
over the underlying graph. In the second model, each node samples its neighbors by following a Markov chain.
The impact of the sampling intervals on consensus convergence and divergence is studied.
We consider consensus in expectation, mean-square, and almost sure sense.
We believe that the models considered in this paper is applicable to some applications since they incorporate sampling by digital devices, limited node connections, and random interactions imposed by unreliable networks.
The main contributions of this paper are summarized as follows.
For both independent and Markovian random network models, necessary and sufficient conditions for mean-square consensus are derived in terms of the spectral radius of the corresponding state transition matrix.
    These conditions can be interpreted as critical thresholds on the inter-sampling interval and we show that
    they can be computed by a generalized eigenvalue problem, which can be stated as a quasi-convex optimization problem.
For each random network model, we  obtain an
upper bound on the inter-sampling interval
below which almost sure convergence is reached, and
 a lower bound on the inter-sampling interval
above which almost sure divergence is reached.
To the best of our knowledge, this
is the first time that almost sure consensus convergence and divergence are
studied for sampled-data systems, and also the first
time that almost sure divergence is considered for Markovian
random graphs.

The remainder of the paper is organized as follows.
Section~\ref{sec:ps} provides the problem formulation, and introduces the probabilistic
consensus notions. Their relations are also discussed. Section~\ref{sec:iid-sampling} focuses on independent
random networks. In this section, we present necessary and/or sufficient conditions for expectation consensus, mean-square consensus, almost sure consensus, and almost sure divergence. The same problems are addressed under a
Markovian network in Section~\ref{sec:markovian-sampling}.
Compared with random networks, Markovian networks
allow each link to be a channel with ``memory''.
In Section~\ref{sec:example}, we illustrate our theoretical results through numerical simulations. Finally, some concluding remarks are drawn in Section~\ref{sec:conclusion}.

\textit{Notations}:
$\mathbb{N}$, $\bb{C}$,~$\mathbb{R}$ and $\mathbb{R}_+$ are the sets of nonnegative integers,
complex numbers, real numbers and positive real numbers, respectively.
For $x,y\in\mathbb{R}$, $x\vee y$ and  $x\wedge y$ stand for
the maximum and minimum of $x$ and $y$ respectively.
The set of $n$ by $n$ positive semi-definite (positive definite) matrices over the field {$\bb{C}$} is denoted as $\mathbb{S}_{+}^{n}$ ($\mathbb{S}_{++}^{n}$).
For a matrix $X=[x_1~x_2~\cdots~x_n]\in
\mathbb{R}^{m\times n}$,
 $\|X\|$ represents the spectral norm of $X$;
$X^*$ and $X'$  are the Hermitian conjugate and the transpose of $X$ respectively.
The Kernel of $X$ is defined as
$\ker(X) = \left\{ {v}\in \bb{R}^{n}: X{v}=0 \right\}.$
$\mathrm{vec}(X)$ is the vectorization of $X$, i.e.,
$\mathrm{vec}(X):=\left[x_1',x_2',\ldots,x_n'\right]'\in\mathbb{R}^{mn}.$
$\otimes$ denotes a Kronecker product of two matrices.
If $m=n$, $\rho(X)$ and $\mathrm{Tr}(X)$ are the spectral radius and
the trace of $X$ respectively.
For vectorization and Kronecker product, the following properties are frequently
used in this work:
\begin{inparaenum}[\itshape i\upshape)]
\item $\mathrm{vec}(ABC)=(C'\otimes A)\mathrm{vec}(B)$;
\item $(A\otimes B)(C\otimes D)=(AC)\otimes (BD),$
\end{inparaenum}
where $A,~B,~C$ and $D$ are matrices of compatible dimensions.
For vectors $x,y\in\bb{R}^{n}$, $x\perp y$ is a short hand
for $\langle x,y\rangle=0$, where $\langle \cdot, \cdot\rangle$
denotes Euclidean inner product.
The indicator function of a subset $\mathscr{A}\subset\Omega$ is a function
$1_\mathscr{A}:\Omega \rightarrow \{ 0,1 \}$, where
$1_\mathscr{A}(\omega)=1$ if $\omega\in \mathscr{A}$, and
$1_\mathscr{A}(\omega)=0$ if $\omega\not\in \mathscr{A}$.
The notation $\sigma(\cdot)$ represents
the $\sigma$-algebra generated by random variables.
Depending on the argument, $|\cdot|$ stands for the absolute value of a real number, or
the cardinality of a set.

\section{Problem Formulation}\label{sec:ps}

\subsection{Sampling and Random Networks}
Consider a network of  $\mathsf{N}$ nodes indexed in the set  $\mathrm{V}=\{1,2,\ldots,\sfN\}$. Each node $i$ holds a value $x_i(t) \in\mathbb{R}$ for $t\in[0,\infty)$. The evolution of $x_i(t)$ is described by
\begin{equation}
\dot x_i(t)=u_i(t),
\end{equation}
where  $u_i\in \mathbb{R}$ is the control input.

The directed interaction graph $\mathrm{G}=(\mathrm{V}, \mathrm{E})$ describes underlying
information exchange. Here $\mathrm{E}\subseteq\mathrm{V}\times\mathrm{V}$ is an arc set and $(j,i)\in\mathrm{E}$ means there is a (possibly unreliable) communication link
 from node
$j$ to node $i$. The set of {neighbors} of node $i$ in the underlying graph $\mathrm{G}$ is
denoted as $\mathscr{N}_i:=
\{j:( j, i)\in \mathrm{E}\}$.
The Laplacian matrix
${L}:=[l_{ij}]\in\mathbb{R}^{\sfN\times \sfN}$ associated
with $\mathrm{G}$ is defined as
\begin{equation*}
{l}_{ij}=\left\{\begin{array}{lll}
-1,& \hbox{if~}i\not=j\hbox{~and~}(j,i)\in\mathrm{E}\\
\sum_{m\not=i}{1}_{\{(m,i)\in\mathrm{E}\}}, &
\hbox{if~}i=j.\end{array}
\right.
\end{equation*}
A directed \emph{path} from node $i_{1}$ to node $i_{l}$ is a sequence of nodes $\{i_1,\ldots,i_l\}$ such that $(i_j,i_{j+1}) \in \mathrm{E}$ for $j=1,\ldots,l-1$. A directed tree is a directed subgraph of  $\mathrm{G}=(\mathrm{V}, \mathrm {E})$
such that every node has exactly one parent, except a single root node with no parents. Therefore,
there must exist a directed path from the root to every other node.
A directed spanning tree is a directed tree that contains all the nodes of $\mathrm{G}$.

Let $\mathscr{G}$ be associated with $\mathrm{G}$ and
the set containing all subgraphs of $\mathrm{G}$ and $\{\mathrm{G}_k=(\mathrm {V},\mathrm
{E}_k)\}_{k\in\mathbb{N}}$ be a sequence of  random graphs, in which by definition each $\mathrm{G}_k$ is  a random variable  taking  values in $\mathscr{G}$.
The Laplacian matrix
${L}(k):=[l_{ij}(k)]\in\mathbb{R}^{\sfN\times \sfN}$ associated
with $\mathrm{G}_k$ is defined as
{\begin{equation*}
{l}_{ij}(k)=\left\{\begin{array}{lll}
-1,& \hbox{if~}i\not=j\hbox{~and~}(j,i)\in\mathrm{E}_k\\
\sum_{m\not=i}{1}_{\{(m,i)\in\mathrm{E}_k\}}, &
\hbox{if~}i=j.\end{array}
\right.
\end{equation*}}
The set of {neighbors} of node $i$ in
denoted as $\mathscr{N}_i(k):=
\{j:( j, i)\in \mathrm{E}_k\}$.
Let the triple $(\mathscr{G}^{\bb{N}},\mathcal{F}, \bb{P})$ denote the probability space capturing the randomness contained in the
random graph sequence, where $\mathcal{F}$ is the set of all subsets of $\mathscr{G}^{\bb{N}}$. Furthermore, we define a filtration
$\mathcal{F}_k=\sigma(\mathrm{G}_0,\ldots, \mathrm{G}_k)$ for $k\in\bb{N}$.

We define a  sequence of node sampling instants
as $0=t_0<\dots<t_k<t_{k+1}<\dots$ with $\tau_k=t_{k+1}-t_k$
representing the inter-sampling interval. The sampled-data consensus scheme associated with the random graph sequence $\{\mathrm{G}_k\}_{k\in\mathbb{N}}$ is given by
\begin{equation}\label{1}
u_i(t)=
\sum_{j\in \mathscr{N}_i(k)}\big[x_j(t_k)-x_i(t_k)\big],\ \ t\in[t_k, t_{k+1}).
\end{equation}
The closed-loop system can then be written in the compact form
\begin{equation}\label{eqn:evolution-xt}
x(t_{k+1})=\big[I-\tau_k{L}(k)\big]x(t_{k}):=W(k)x(t_{k})
\end{equation}
with $W(k):=[w_{ij}(k)]$.
\begin{remark}
In the sampled-data algorithm \eqref{eqn:evolution-xt}, each node samples its own state at the sampling instants $\{t_k\}_{k=0}^\infty$. If each node has continuous access to its own state for all $t\geq 0$, we can introduce the algorithm
\begin{equation}\label{2}
u_i(t)=\sum_{j\in \mathscr{N}_i(t_k)} [x_j(t_k)-x_i(t)],\ \   t\in[t_k, t_{k+1}),
\end{equation}
as considered in~\cite{FengLong_TAC08}.
The corresponding closed-loop system is then
\begin{equation}\label{eqn:model-neighboring-sampling}
x(t_{k+1})=\big[I-(1-e^{-\tau_k}){L}(k)\big]x(t_{k}).
\end{equation}
By replacing $\tau_k$ in~\eqref{eqn:evolution-xt} with
$1-e^{-\tau_k}$ in~\eqref{eqn:model-neighboring-sampling},
all the conclusions in this paper for~\eqref{eqn:evolution-xt} throughout
the paper can thus be readily translated into those for~\eqref{2}.
\end{remark}
\subsection{Consensus Metrics}
Define
$x_{\max}(t_k):= \max_{i\in \mathrm{V}} x_i(t_k)$ and  $x_{\min}(t_k):= \min_{i\in \mathrm{V}} x_i(t_k)$
and the agreement measure
$\mathfrak{X}(k):= x_{\max}(t_k)-x_{\min}(t_k).$
We have the following definitions
for consensus convergence and divergence.
\begin{definition}\label{def:expect-convergence}
\begin{itemize}

\item[(i)] Algorithm \eqref{eqn:evolution-xt}  achieves (global) consensus in expectation if
for any initial state $x(t_0)\in \mathbb{R}^{\mathsf{N}}$  there holds $\lim_{k\rightarrow \infty}\mathbb{E}[\mathfrak{X}(k)]=0.$

\item[(ii)] Algorithm \eqref{eqn:evolution-xt}  achieves (global)  consensus in mean square if for any initial state $x(t_0)\in \mathbb{R}^{\mathsf{N}}$ there holds  $\lim_{k\rightarrow \infty}\mathbb{E}[\mathfrak{X}^2(k)]=0.$

\item[(iii)] Algorithm \eqref{eqn:evolution-xt}  achieves (global)  consensus almost surely if for any initial state $x(t_0)\in \mathbb{R}^{\mathsf{N}}$ there holds
$\mathbb{P}\left(\lim_{k\rightarrow \infty}\mathfrak{X}(k)=0\right)=1.$

\item[(iv)] Algorithm \eqref{eqn:evolution-xt}  diverges
almost surely
if there holds $\mathbb{P}\Big(\lim\sup_{k\rightarrow \infty}\mathfrak{X}(k)=\infty\Big)=1$ for any initial state $x(t_0)\in\mathbb{R}^{\sfN}$ except for $x(t_0)\perp \mathbf{1}$.
\end{itemize}
\end{definition}

\subsection{Relations of Consensus Notions}
The following lemma suggests that if the inter-sampling interval is small enough, the consensus notations in Definition~\ref{def:expect-convergence} are equivalent.
\begin{lemma}\label{lemma:equ-consensus-notaions}
Suppose $\tau_k\in\left(0,(\mathsf{N}-1)^{-1}\right]$ for all $k$. Then expectation consensus, mean-square consensus, and almost sure consensus  are all equivalent for  Algorithm \eqref{eqn:evolution-xt}.
\end{lemma}
\textit{Proof.}
We begin with the observation that $W(k)$ is a
row stochastic matrix for all $k\in\mathbb{N}$ when
$\tau_k\in\left(0,(\mathsf{N}-1)^{-1}\right]$,
where a row stochastic matrix means a nonnegative square matrix with
each row summing to $1$.
Therefore,
$$x_{\max}(t_{k+1})=
\max_{i\in\mathrm{V}}\sum_{j=1}^\mathsf{N}w_{ij}(k)x_j(t_k)
\leq \max_{i\in\mathrm{V}}\sum_{j=1}^\mathsf{N}w_{ij}(k)
\big(x_j(t_k)\vee x_{\max}(t_k)\big)=x_{\max}(t_k),$$
implying that $x_{\max}(t_k)$ is non-increasing in $k$.
We show that $x_{\min}(t_k)$ is non-decreasing in $k$ in
precisely the same way. The foregoing two observations together suggest that $\mathfrak{X}(k)$ is non-increasing in $k$.
Finally, the conclusion follows by
showing the following implications:
\begin{enumerate}

\item[(i)]
Expectation consensus $\Longrightarrow$ mean-square consensus.
~Since $\mathfrak{X}(k)$ is non-increasing, we have $\mathbb{E}[\mathfrak{X}^2(k)]\leq \mathfrak{X}(0)\mathbb{E}[\mathfrak{X}(t)]$.
By the hypothesis, $\mathbb{E}[\mathfrak{X}^2(k)]\leq \mathfrak{X}(0)\mathbb{E}[\mathfrak{X}(k)]\to 0$ as $k\to\infty$.

\item[(ii)]
Mean-square consensus $\Longrightarrow$ almost sure consensus.
According to Chebyshev's inequality~\cite{hardy1952inequalities},
$$
\mathbb{P}\left(|\mathfrak{X}(k)|<\epsilon\right)\leq \frac{\mathbb{E}[\mathfrak{X}^2(k)]}{\epsilon^2}
$$
holds for any $\epsilon>0$.
If $\lim_{k\rightarrow \infty}\mathbb{E}[\mathfrak{X}^2(k)]=0$, consequently $\lim_{k\rightarrow \infty}\mathbb{P}\left(|\mathfrak{X}(k)|<\epsilon\right)=0$. As a result, there exists
a subsequence of $\{\mathfrak{X}(k)\}_{k\in\mathbb{N}}$ that converges to $0$
almost surely~\cite{durrett2010probability}. Since $\{\mathfrak{X}(k)\}_{k\in\mathbb{N}}$ is non-increasing, $\lim_{k\rightarrow \infty}\mathfrak{X}(k)= 0$ almost surely.

\item[(iii)]
Almost sure consensus $\Longrightarrow$ expectation consensus.
Since the sequence $\{\mathfrak{X}(k)\}_{k\in\mathbb{N}}$ is nonnegative and non-increasing, by the Monotone Convergence Theorem~\cite{durrett2010probability}, $\lim_{k\rightarrow \infty}\mathbb{E}[\mathfrak{X}(k)]=0$.
\end{enumerate}
\hfill $\square$
\begin{remark}
In~\cite{Guanrong11tac}, the equivalence of $L^p$ consensus,
consensus in probability, and almost sure consensus was obtained
over a random network
generated by i.i.d. stochastic matrices.
In Lemma~\ref{lemma:equ-consensus-notaions}, we show that
this equivalence holds regardless of the type of random process the row stochastic matrices
are generated by.
 The equivalence relation follows from the
monotonicity of $\{\mathfrak{X}(k)\}_{k\in\mathbb{N}}$.
\end{remark}

\section{Independent  Random  Networks}\label{sec:iid-sampling}
In this section, we investigate sampled-data consensus when the random graph $\mathrm{G}_k$ is obtained by each node independently sampling its neighbors in a random manner over $\mathrm{G}$.
Regarding the connectivity of the underlying graph $\mathrm{G}$, we adopt the following assumption:
    \begin{enumerate}
    \item[(A1)]\label{asmpt:assumpt-spanning-tree}
    \emph{The underlying graph $\mathrm{G}$
    has a directed spanning tree.}
    \end{enumerate}
    We also impose the following assumption.
    \begin{enumerate}
    \item[(A2)]\label{asmpt:assumpt-independent-sampling}
     \emph{The random variables ${1}_{\{(j,i)\in
    \mathrm{E}_k\}}$, $(j,i)\in \mathrm{E}$, $k\in\bb{N}$, are
    i.i.d. Bernoulli with mean $q>0$.}
    \end{enumerate}
The techniques developed
in this section also apply when $q=q(i)$ is a function of
node index $i$.

In order to simplify the notation used in the derivation of
the results through this section, we also make
the following assumption.
\begin{enumerate}
\item[(A3)] Let $\tau_k=\tau_\ast$ for all $k$ with $\tau_\ast>0$.
\end{enumerate}

When each node samples its neighbors as Assumption~(A2) describes,
$\{{L}(k)\}_{k\in\bb{N}}$ are i.i.d. random variables, whose
 randomness originates from the primitive random variables ${1}_{\{(j,i)\in \mathrm{E}_k\}}$'s. We denote the sample space of ${L}(k)$ by $\mathscr{L}:=\{{L}^{(1)},{L}^{(2)},\ldots,{L}^{(\sfM)}\}$ where
 $\sfM=|\mathscr{G}|$ and $L^{(l)}:=\big[l_{ij}^{(l)}\big]\in\bb{R}^{\sfN\times \sfN}$ is the
 Laplacian matrix associated with a subgraph $\mathrm{G}^{(l)}\in\mathscr{G}$.
 By counting how many edges are present in $\mathrm{G}_k$
 and how many are absent from $\mathrm{G}_k$, respectively,
the distribution of ${L}(k)$ is computed by
\begin{equation}\label{eqn:iid-prob}
\mathbb{P}({L}(k)={L}^{(i)})=
q^{\mathrm{Tr}({L}^{(i)})}(1-q)^{\mathrm{Tr}(L-{L}^{(i)})}:=
\pi_i,~~~i=1,\ldots,\sfM.
\end{equation}
When $\tau_k=\tau_*$,
$W(k)$ inherits the same distribution as ${L}(k)$ from $\mathrm{G}_k$.
Then, we denote ${W}^{(l)}=:I-\tau_*{L}^{(l)}$.

\subsection{Conjunction of Various Consensus Metrics}
When the inter-sampling interval is small enough (to be precise， $\tau_*<(\mathsf{N}-1)^{-1}$), each node recursively updates its state
as a convex combination of the previous states of its own and its neighbors.
Every update drives nodes' states closer to each other
and can be thought of as attraction of the nodes' states.
Under the independent random network model,
we show in the following theorem that,
as long as $\mathrm{G}$ has a directed spanning tree,
Algorithm \eqref{eqn:evolution-xt} achieves consensus,
simultaneously in expectation,
in mean square, and in almost sure sense.
\begin{theorem}\label{thm:thm1}
Let Assumptions~(A1), (A2), and (A3) hold. Then expectation consensus, mean-square consensus, and almost sure consensus are
 achieved under Algorithm \eqref{eqn:evolution-xt} if { $\tau_\ast\in\big(0,(\mathsf{N}-1)^{-1}\big)$.}
\end{theorem}
\textit{Proof.}
By Lemma~\ref{lemma:equ-consensus-notaions}, it suffices to show that Algorithm~\eqref{eqn:evolution-xt} achieves consensus in expectation.

Fix a directed spanning tree $\mathrm{G}_T$ of
 graph $\mathrm{G}$ and a sampling time $t_k$.
Let the root of $\mathrm{G}_T$ be $i_{1}\in\mathrm{V}$, and define
a set of nodes $\mathscr{M}_1:=\{i_1\}$.
Denote $$\eta:=( \tau_\ast)\wedge(1-(\sfN-1) \tau_\ast).$$
Then, there holds $\eta>0$ when $\tau_\ast\in\big(0,(\mathsf{N}-1)^{-1}\big)$.
We first assume $x_{i_{1}}(t_k)\leq 1/2(x_{\max}(t_k)+x_{\min}(t_k))$ while the
other case for $x_{i_{1}}(t_k)> 1/2(x_{\max}(t_k)+x_{\min}(t_k))$
will be discussed later.

Choose a node $i_2\in\mathrm{V}$ such that
$i_2\not\in\mathscr{M}_1$ and
$(i_1,i_2)\in\mathrm{G}_T$. Define $\mathscr{M}_2:=
\mathscr{M}_1\cup\{i_2\}$.
Consider the event
$\mathscr{E}_2:=\left
\{(i_1,i_2)\in \mathrm{E}_{k+1}\right\}$.
When $\mathscr{E}_2$ happens, $x_{i_{2}}(t_{k+1})$ evolves as follows:
\begin{align*}
x_{i_{2}}(t_{k+1})&= w_{i_2i_1}(k)x_{i_1}(t_k)+\sum_{j\not=i_1}
w_{i_2j}(k)x_j(t_k)\\
&\leq \frac{1}{2}w_{i_2i_1}(k)(x_{\min}(t_k)+x_{\max}(t_k))+
(1-w_{i_2i_1}(k))x_{\max}(t_k)\\
&\leq \frac{1}{2}\eta x_{\min}(t_k)+(1-\frac{1}{2}\eta)
x_{\max}(t_k),
\end{align*}
where the last inequality holds because
$\eta\leq w_{i_2i_1}(k)$.
Since $\eta\leq w_{i_1i_1}(k)$, we show that $x_{i_1}(t_{k+1})$ is bounded by
$$
x_{i_1}(t_{k+1})\leq\frac{1}{2}\eta x_{\min}(t_k)+(1-\frac{1}{2}\eta)
x_{\max}(t_k).
$$
At time $t_{k+2}$,
\begin{align*}
x_{i_2}(t_{k+2})&=
w_{i_2i_2}(k+1)x_{i_2}(t_{k+1})+
\sum_{j\neq i_1} w_{i_2j}(k+1)x_j(t_{k+1})\\
&\leq  w_{i_2i_2}(k+1)\left[\frac{1}{2}\eta x_{\min}(t_{k})+(1-\frac{1}{2}\eta)x_{\max}(t_{k})\right]+
\big(1-w_{i_2i_2}(k+1)\big)x_{\max}(t_{k+1})\\
&\leq  \frac{1}{2} \eta^2 x_{\min}(t_k)+
(1-\frac{1}{2}\eta^2)x_{\max}(t_k),
\end{align*}
where the last inequality is
due to $x_{\max}(t_{k+1})\leq x_{\max}(t_k)$ and
$\eta\leq w_{i_2i_2}(k+1)$.
The same is true of node $i_1$, i.e.,
$ x_{i_1}(t_{k+2})\leq \frac{1}{2} \eta^2 x_{\min}(t_k)+(1-\eta^2)x_{\max}(t_k)$.
Recursively, we see that
$$x_{i_1}(t_{k+n})\leq \frac{1}{2} \eta^n x_{\min}(t_k)
+(1-\frac{1}{2}\eta^n)x_{\max}(t_k)$$
and
$$
x_{i_2}(t_{k+n})\leq \frac{1}{2}
\eta^n x_{\min}(t_k)+(1-\frac{1}{2}\eta^n)x_{\max}(t_k).
$$
holds for $n=1,2,\ldots$.

Again, choose a node $i_3\in\mathrm{V}$ such that
$i_3\not\in\mathscr{M}_2$ and there exists
a node $j\in\mathscr{M}_2$ satisfying
$(j,i_3)\in\mathrm{G}_T$. Define $\mathscr{M}_3:=
\mathscr{M}_2\cup\{i_3\}$.
Consider the event
$\mathscr{E}_3:=\left
\{(j,i_3)\in \mathrm{E}_{k+2}:(j,i_3)\in\mathrm{G}_T,j\in\mathscr{M}_2\right\}.$
If $\mathscr{E}_3$ happens, we obtain a similar result for node $i_3$:
\begin{align*}
x_{i_3}(t_{k+2})&\leq
\eta\left(x_{i_1}(t_{k+1})\vee x_{i_2}(t_{k + 1})\right)+
(1-\eta)x_{\max}(t_{k+1})\\
&\leq  \frac{1}{2} \eta^2 x_{\min}
(t_k)+(\eta-\frac{1}{2}\eta^2)x_{\max}(t_k)+(1-\eta)x_{\max}(t_{k})\\
&=\frac{1}{2} \eta^2 x_{\min}
(t_k)+(1-\frac{1}{2}\eta^2)x_{\max}(t_k).
\end{align*}
From the same argument as above,
$$x_{i_3}(t_{k+n})\leq \frac{1}{2} \eta^n x_{\min}(t_k)+(1-\frac{1}{2}\eta^n)x_{\max}(t_k)$$
holds for $n=2,3,\ldots$.

We choose nodes $i_1,\ldots,i_{\mathsf{N}}$ in
sequel and accordingly define $\mathscr{M}_1,\ldots,\mathscr{M}_{\mathsf{N}}$
and $\mathscr{E}_2,\ldots,\mathscr{E}_{\mathsf{N}}$. Consider $\mathscr{E}_2,\ldots,\mathscr{E}_{\mathsf{N}}$
sequentially happen, then
$$x_{i_m}(t_{k+n})\leq \frac{1}{2} \eta^{n}
x_{\min}(t_k)+(1-\frac{1}{2}\eta^{n})x_{\max}(t_k)
$$
holds for all $1\leq m\leq \sfN$ and $n\geq \mathsf{N}-1$, which entails
$$x_{\max}(t_{k+\sfN-1})=\max_i x_i(t_{k+\sfN-1})
\leq \frac{1}{2} \eta^{\sfN-1} x_{\min}(t_k)+(1-\frac{1}{2}\eta^{\sfN-1})x_{\max}(t_k).$$
In this case, the relationship between $\mathfrak{X}(t_{k+\mathsf{N}-1})$ and
$\mathfrak{X}(k)$ is given by
\begin{align*}
\mathfrak{X}({k+\mathsf{N}-1})&=x_{\max}
(t_{k+\mathsf{N}-1})-x_{\min}(t_{k+\mathsf{N}-1})\\
&\leq \frac{1}{2} \eta^{\mathsf{N}-1} x_{\min}(t_k)+(1-\frac{1}{2}\eta^{\mathsf{N}-1})x_{\max}(t_k)-x_{\min}(t_k)\\
&=\left(1-\frac{1}{2}\eta^{\mathsf{N}-1}\right)\mathfrak{X}(k).
\end{align*}
If $x_{i_1}(t_k)> 1/2\left(x_{\max}(t_k)+x_{\min}(t_k)\right)$ is assumed, a symmetric analysis leads to that, when $\mathscr{E}_2,\ldots,\mathscr{E}_{\mathsf{N}}$
sequentially occur,
$x_{\min}(t_{k+\sfN-1})\geq \frac{1}{2}\eta^{\sfN-1}x_{\max}(t_k)+
(1-\frac{1}{2}\eta^{\sfN-1})x_{\min}(t_k)$. Then $\mathfrak{X}({k+\mathsf{N}-1})$ is
bounded by
\begin{align*}
\mathfrak{X}({k+\mathsf{N}-1})&=x_{\max}(t_{k+\mathsf{N}
-1})-x_{\min}(t_{k+\mathsf{N}-1})\\
&\leq x_{\max}(t_k)-
\frac{1}{2} \eta^{\mathsf{N}-1}
x_{\max}(t_k)-(1-\frac{1}{2}\eta^{\mathsf{N}-1})x_{\min}(t_k)\\
&=\left(1-\frac{1}{2}\eta^{\mathsf{N}-1}\right)\mathfrak{X}(k),
\end{align*}
exactly the same result as when
 $x_{i_{1}}(t_k)\leq 1/2(x_{\max}(t_k)+x_{\min}(t_k))$ is assumed. Therefore, the above inequality holds irrespective of the state of $x_{i_{1}}(t_k)$.

In addition, we know that probability that the events $\mathscr{E}_2,\ldots,\mathscr{E}_{\mathsf{N}}$
sequentially occur is
$$\mathbb{P}\left({1}_{\cap_{i=2}^{\mathsf{N}}
\mathscr{E}_i}=1\right)=
\prod_{i=2}^{\mathsf{N}}\mathbb{P}({1}_{\mathscr{E}_i}=1)\geq
q^{\mathsf{N}-1}.$$
Combining all the above analysis,
\begin{align}\label{eqn:coverge-rate}
\mathbb{E}[\mathfrak{X}({k+\mathsf{N}-1})]&\leq
q^{\mathsf{N}-1}\left(1-\frac{1}{2}\eta^{\mathsf{N}
-1}\right)\mathbb{E}[\mathfrak{X}(k)]
+(1-q^{\mathsf{N}-1})\mathbb{E}[\mathfrak{X}(k)]\notag\\
&=\left(1-\frac{1}{2}(q \eta)^{\mathsf{N}-1}\right)\mathbb{E}[\mathfrak{X}(k)].
\end{align}
Since $0<q\eta<1$, then $\lim_{k\to\infty}\mathbb{E}[\mathfrak{X}(k)]=0$, which
 completes the proof.
\hfill $\square$

 When the inter-sampling interval $\tau_\ast$ is too large,
 then $W(k)$ may have negative entries. Consequently, some
 nodes may repel, so consensus of
 Algorithm~\eqref{eqn:evolution-xt} may not be achieved.
When repulsive actions exist,
expectation consensus, mean-square consensus, and almost sure consensus
are not equivalent in general since the Monotone
Convergence Theorem
cannot be applied. Of course, consensus in mean square
still implies expectation consensus as consistent
with ${L}^p$ convergence for a sequence of random variables.
In the subsequent two subsections, mean-square consensus and
almost sure consensus/divergence will be separately analyzed.

\subsection{The Mean-square Consensus Threshold}

In this part, we focus on mean-square consensus. First of all, we give
a necessary and sufficient mean-square consensus condition in terms of the spectral radius of a matrix that depends on $\tau_\ast$, $\mathrm{G}$ and $q$, by
studying the spectral property of a linear system. Note that the analysis is carried out on the spectrum restricted to the
 smallest invariant subspace containing
$I-\frac{1}{\mathsf{N}}\mathbf{1}\mathbf{1}'$.
The condition is then interpreted as the existence of a critical threshold on the inter-sampling intervals,
 below which
Algorithm~\eqref{eqn:evolution-xt} achieves mean-square consensus
and above $\mathfrak{X}(k)$ diverges
in mean-square sense for some initial state $x(t_0)$.
This translation relies on the relationship between
the stability of a certain matrix and
the feasibility of a linear matrix inequality.

\begin{proposition}\label{thm:iff-mean-square-consensus}
Let Assumptions~(A1),~(A2), and (A3) hold. Then
the following statements are equivalent:
\begin{itemize}
\item[(i)] Algorithm \eqref{eqn:evolution-xt} achieves mean-square consensus;
\item[(ii)] There holds $\rho\Big(\mathbb{E}[W(0)\otimes W(0)](J\otimes J)\Big)<1$, where \begin{equation}\label{def:J-matrix}
J:=I-\frac{1}{\mathsf{N}}\mathbf{1}\mathbf{1}';
\end{equation}
\item[(iii)] There exists a matrix $S>0$ such that
\begin{equation}\label{def:phi}
\phi(S):=\sum_{i=1}^{\sfM}\pi_i
J{W}^{(i)}JSJ({W}^{(i)})'J<S,
\end{equation}
where $\pi_i$ is defined in~\eqref{eqn:iid-prob}.
\end{itemize}
\end{proposition}
\textit{Proof.}
The proof needs the following lemma.
\begin{lemma}[Lemma~2~in~\cite{Costa04TAC}]\label{lemma:matrix-depomositation}
For any $G\in \mathbb{C}^{n\times n}$ there exist $G_i\in
\mathbb{S}^{n}_+,~i=1,2,3,4$, such that
$$G=(G_1-G_2)+(G_3-G_4)\mathbf{i}$$
where $\mathbf{i}=\sqrt{-1}$.
\end{lemma}

Define the difference between the state $x(t_k)$ and
its average as
\begin{equation}\label{eqn:def-dk}
d(k):=x(t_k)-\frac{1}{\mathsf{N}}\mathbf{1}\mathbf{1}'x(t_k).
\end{equation}
Evidently, $d(k)=Jx(t_k)$. Since
\begin{align}\label{eqn:dt-lower-bound}
\mathfrak{X}(k)&=
x_{\max}(k)-\frac{1}{\mathsf{N}}\mathbf{1}'x(t_k)
-\left[x_{\min}(t_k)-\frac{1}{\mathsf{N}}\mathbf{1}'x(t_k)\right]\notag\\
&\leq\left|x_{\max}(t_k)-\frac{1}{\mathsf{N}}\mathbf{1}'x(t_k)\right|
+\left|x_{\min}(t_k)-\frac{1}{\mathsf{N}}\mathbf{1}'x(t_k)\right|\notag\\
&\leq\sqrt{2\sum_{i=1}^\mathsf{N}\left[x_i(t_k)-
\frac{1}{\mathsf{N}}\mathbf{1}'x(t_k)
\right]^2}\notag\\
&=\sqrt{2}\,\|d(k)\|
\end{align}
and
\begin{equation}\label{eqn:dt-upper-bound}
\mathfrak{X}(k)=
{\mathsf{N}^{-1/2}}\sqrt{\sfN (x_{\max}(t_k)-x_{\min}(t_k))^2}
\geq
{\mathsf{N}^{-1/2}}\sqrt{\sum_{i=1}^\mathsf{N}
\left[x_i(t_k)-\frac{1}{\sfN}\mathbf{1}'x(t_k)
\right]^2}={\mathsf{N}^{-1/2}}\,\|d(k)\|,
\end{equation}
$\lim_{k\to\infty}\mathbb{E}[\mathfrak{X}^2(k)]=0$ is equivalent to
$\lim_{k\to\infty}\mathbb{E}\|d(k)\|^2=0$.
From the Cauchy-Schwarz inequality, $\mathbb{E}[d_i(k)d_j(k)]\leq
\mathbb{E}[d_i(k)^2]^{1/2} \mathbb{E}[d_j(k)^2]^{1/2}$ holds for any $1\leq i,j\leq \mathsf{N}$, which furthermore implies the equivalence between
$\lim_{k\to\infty}\mathbb{E}\|d(k)\|^2=0$ and  $\lim_{k\to\infty}\mathbb{E}[d(k)d(k)^*]=0$. Thus, to study the mean-square consensus, we only need to focus on whether  $\mathbb{E}[d(k)d(k)^*]$ converges to a zero matrix.

Observe that
 \begin{align}
 d({k}) &=JW(k-1)x(t_{k-1})\notag\\
 &=JW(k-1)x(t_{k-1})-\frac{1}{\mathsf N}JW(k-1)\mathbf{1}\mathbf{1}'x(t_{k-1})\notag\\
 &=JW(k-1)d({k-1}) \label{eqn:evolution-d_kt}
 \end{align}
 holds for $k=1,2,\ldots$, where the second equality is due to
 $JW(k)\mathbf{1}=J\mathbf{1}=0$. It entails
 $$\mathbb{E}[d(k)d(k)^*]=\mathbb E\big[JW(k-1)d(k-1)d(k-1)^*W(k-1)'J\big].$$
Taking vectorization on both sides  yields
\begin{align}\label{eqn:iteration-dk}
\mathrm{vec}\big(\mathbb{E}[d(k)d(k)^*]\big)
&= \mathbb{E}\left[(JW(k-1))\otimes(JW(k-1))\mathrm{vec}\big(d(k-1)d(k-1)^*\big)
\right]\notag\\
&= \mathbb{E}\left[(JW(0))\otimes(JW(0))\right] \mathrm{vec}\left(\mathbb{E}[d(k-1)d(k-1)^*]\right)\notag\\
&=( J\otimes J)\mathbb{E}[W(0)\otimes W(0)]\, \mathrm{vec}\left(\mathbb{E}[d(k-1)d(k-1)^*]\right)\notag\\
&=
\Big(( J\otimes J)\mathbb{E}[W(0)\otimes W(0)]\Big)^k \mathrm{vec}\left(d(0)d(0)^*\right)\notag\\
&=\Big(( J\otimes J)\mathbb{E}[W(0)\otimes W(0)]\Big)^k ( J\otimes J)\,
\mathrm{vec}\left(x(t_0)x(t_0)^*\right)\notag\\
&=( J\otimes J)\Big(\mathbb{E}[W(0)\otimes W(0)]( J\otimes J)\Big)^k
\mathrm{vec}\left(x(t_0)x(t_0)^*\right),
\end{align}
where the first equality is based on the property $\mathrm{vec}(ABC)=(C'\otimes A)\mathrm{vec}(B)$ for matrices $A,~B$ and $C$ of compatible dimensions, and
the separation of expectations in the second equality is due to the independence of the random interconnections.

The implications from one statement to the next is provided as follows.

\noindent $(i)\Rightarrow (ii)$. If
$\rho\Big(\mathbb{E}[W(0)\otimes W(0)] (J\otimes J)\Big)\geq1$,
there exist a number
$\lambda$ with $|\lambda|\geq 1$ and a non-zero vector $v\in\mathbb{C}^{\mathsf{N}^2}$
corresponding to $\lambda$ satisfying
$\mathbb{E}[W(0)\otimes W(0)]( J\otimes J)v=
\lambda v$.
Let $v_1,\ldots,v_l$ be all the eigenvectors
corresponding to the eigenvalue $0$ of $J\otimes J$.
Since $\mathbb{E}[W(0)\otimes W(0)]( J\otimes J)v_i=0$ for
any $i=1,\ldots,l$,
 there holds $v\not=\sum_{i=1}^la_iv_i$ for any
 $a_i\in\mathbb{R}$ and
 $(J\otimes J)v\not =0$. Therefore
\begin{equation}\label{eqn:intial-v-infty}
\lim_{k\to\infty}( J\otimes J)
\Big(\mathbb{E}[W(0)\otimes W(0)]( J\otimes J)\Big)^k
v=\lim_{k\to\infty}\lambda^k( J\otimes J)v\not =0.
\end{equation}
In order to show that mean-square consensus is not achieved for Algorithm \eqref{eqn:evolution-xt}, it remains to prove that $v$ can be expressed as
a linear combination of different initial states.
Note that there exist $G\in\mathbb{C}^{\mathsf{N\times N}}$ and
$G_1,\ldots,G_4\in\mathbb{S}_+^{\mathsf{N}}$ such that $v=\vec(G)$
and $G=G_2-G_4+(G_3-G_1)\mathbf{i}$  by Lemma~\ref{lemma:matrix-depomositation} (the order of
$G_1,~G_2,~G_3$ and $G_4$ is immaterial in this lemma).
Since each $G_i$ can be expressed as
$$G_i=\sum_{j=1}^{\mathsf{N}}\lambda_{j}^{(i)} u_{j}^{(i)}(u_{j}^{(i)})^*,$$
where $G_i=U^{(i)}\mathrm{diag}\{\lambda_{1}^{(i)},\ldots,
\lambda_{\mathsf{N}}^{(i)}\}(U^{(i)})^*$ with
$\lambda_{j}^{(i)}\in\sigma(G_i)$ and
$U^{(i)}=:[u_{1}^{(i)},\ldots,u_{\mathsf{N}}^{(i)}]$
unitary.
Then,
we have
\begin{equation*}
v=\sum_{i=1}^4\sum_{j=1}^{\mathsf{N}}-\lambda_{j}^{(i)}
{\mathbf{i}}^i
\mathrm{vec}\big(u_{j}^{(i)}(u_{j}^{(i)})^*\big).
\end{equation*}
By letting $x(t_0)=u_{j}^{(i)},~i=1,\ldots,4$ and $j=1,\ldots,\mathsf{N}$, respectively, we see from~\eqref{eqn:intial-v-infty} that mean-square consensus is not
achieved for some $x(t_0)$.

\noindent $(ii)\Rightarrow (iii)$.
Denote $R:=(J\otimes J)\mathbb{E}[W(0)\otimes W(0)] (J\otimes J)$.
From $(ii)$,
\begin{align*}
\rho(R)&=\rho\Big(\mathbb{E}[W(0)\otimes W(0)] (J\otimes J)^2\Big)\\
&=\rho\Big(\mathbb{E}[W(0)\otimes W(0)] (J\otimes J)\Big)<1.
\end{align*}
Then,
$(I-R)^{-1}$ exists and is nonsingular, $(I-R)^{-1}=
\sum_{j=0}^{\infty}R^j$.
For any given positive definite matrix $V\in\mathbb{R}^{\sfN \times \sfN}$, there corresponds a
unique matrix $S\in\mathbb{R}^{\sfN \times \sfN}$ such that
\begin{equation}\label{eqn:v-p-Lk}
\mathrm{vec}(V)=\left(I-R\right)\mathrm{vec}(S).
\end{equation}
Then,
\begin{align*}
\mathrm{vec}(V)&=\Big(I-\mathbb{E}[(JW(0)J)\otimes (JW(0)J)] \Big)
\mathrm{vec}(S)\\
&=\vec\big(S-\phi(S)\big),
\end{align*}
where $\phi(\cdot)$ is defined in~\eqref{def:phi}, which implies $S-\phi(S)>0$ by the one-to-one correspondence
of the vectorization operator.
The positive definiteness of $S$ follows from
\begin{align}\label{eqn:v-p-Lk}
\mathrm{vec}(S)&=\left(I-R\right)^{-1}\mathrm{vec}(V)\notag\\
&=\sum_{i=0}^{\infty}R^i\,\mathrm{vec}(V)\notag\\
&=\mathrm{vec}\left(\sum_{i=0}^\infty \phi^i(V)\right),\notag
\end{align}
implying
$S=\sum_{i=0}^\infty \phi^i(V)\geq V>0$, again by
the one-to-one correspondence
of the vectorization operator.

\noindent  $(iii)\Rightarrow (i)$.
By the hypothesis, there always exists a $\mu \in (0,1)$ satisfying
$\phi(S)<\mu S$. Fix any given $X\in\mathbb{S}_{+}^{\sfN}$
and then choose a $c>0$ satisfying
$X\leq c S$.
Then, by the linearity and non-decreasing properties
of $\phi(X)$ in $X$ over the positive
semi-definite cone,
$$
\phi^k(X)\leq \phi^k(cS)=
c\phi^k(S)
<c\phi^{k-1}(\mu S)=c\mu\phi^{k-1}(S)
<\cdots< c \mu^k S
$$
holds for all $k\in\bb{N}$. It leads to $\lim_{k \rightarrow \infty}\phi^k(X)=0$, which
 means
\begin{equation}\label{eqn:limit-zero}
\lim_{k\rightarrow \infty} R^k\mathrm{vec}(X)=0.
\end{equation}
In light of Lemma~\ref{lemma:matrix-depomositation}, for any
$G\in \mathbb{R}^{n\times n}$
there exist $X_1,~X_2,~X_3,~X_4\in \mathbb{S}_+^n$
such that $G=(X_1-X_2)+(X_3-X_4)\mathbf{i}$.
Then, we see from~\eqref{eqn:limit-zero}
\begin{align*}
&\lim_{k\rightarrow \infty}R^k\mathrm{vec}(G)\\
=&
\lim_{k\rightarrow \infty} R^k\Big(\mathrm{vec}(X_1)-\mathrm{vec}(X_2)
+\mathrm{vec}(X_3)\mathbf{i}-\mathrm{vec}(X_4)\mathbf{i}\Big)\\
=&0.
\end{align*}
Since $G$ is arbitrarily chosen, we have $\rho\Big(\mathbb{E}[W(0)\otimes W(0)](J\otimes J)\Big)=\rho(R)<1$. Then,
$$\lim_{k\to \infty}\mathrm{vec}\big(\mathbb{E}[d(k)d(k)^*]\big)=( J\otimes J)\lim_{k\to \infty}
\Big(\mathbb{E}[W(0)\otimes W(0)]( J\otimes J)\Big)^k
\mathrm{vec}(x(t_0)x(t_0)^*)=0$$
holds for any $x(t_0)\in\mathbb{R}^{\mathsf{N}}$, which
means $\lim_{k\to \infty}\mathbb{E}[d(k)d(k)^*]=0$.
\hfill $\square$

The following result holds.
\begin{theorem}\label{prop:propostion1}
    Let Assumptions~(A1), (A2), and (A3) hold.
    Then Algorithm \eqref{eqn:evolution-xt} achieves mean-square
    consensus if and only if $\tau_*\leq \tau_\dagger$,
    where
    $\tau_\dagger$ is given by the following quasi-convex optimization problem:
    {
    \begin{align}
    \mathrm{\arg\min}_\tau\, & -\tau \notag\\
    \mathrm{subject~to}&\left[\begin{array}{cccc}
    JZJ+\mathbf{11}'& \sqrt{\pi_1} \left(JZ- JL^{(1)}J Y\right) & \ldots
    &\sqrt{\pi_\sfM}(JZ- J
    {L}^{(\sfM)}J Y)\\
    * & Z &\ldots& 0\\
    \vdots & \vdots & \ddots &\vdots\\
    * &* & \ldots &Z \end{array}\right]>0,\notag\\
    &&\label{eqn:LMI-1}\\
    &Y,Z>0,\label{eqn:LMI-2}\\
    &Y- \tau Z\geq 0,\label{eqn:LMI-3}
    \end{align}}
    with $\ast$'s standing for entries that are the Hermitian conjugates of entries in the upper triangular part.
\end{theorem}
\textit{Proof.}
\noindent  \textit{Necessity:}
Suppose that mean-square consensus is achievable
for Algorithm \eqref{eqn:evolution-xt}, or equivalently
there exists a matrix $S>0$ such that
$\phi(S)<S$ holds by Proposition~\ref{thm:iff-mean-square-consensus}.
 First we shall show $\sum_{i=1}^{\sfM}\pi_i
J{W}^{(i)}JSJ({W}^{(i)})'J<JSJ+\mathbf{11}'$.
Without loss of generality, choose for $(v_1,\ldots, v_{\sfN})$ an orthonormal basis
of $\bb{R}^{\sfN}$ with $v_{1}=\frac{1}{\sfN}\mathbf{1}$. Then, any
vector $0\not=x\in\bb{R}^n$ can be expressed as $x=\sum_{i=1}^{\sfN}a_iv_i$
with coefficients $a_1,\ldots, a_{\sfN}$ not all $0$.
We have
$$x'\phi(S)x=\Big(\sum_{i=2}^{\sfN}a_iv_i\Big)'\phi(S)\Big(\sum_{i=2}^{\sfN}a_iv_i\Big)$$
and
\begin{equation*}
x'(JSJ+ \mathbf{11}')x=
\Big(\sum_{i=2}^{\sfN}a_iv_i\Big)'S\Big(\sum_{i=2}^{\sfN}a_iv_i\Big)
+ a_1^2.
\end{equation*}
Since $a_1,\ldots, a_{\sfN}$ are not all $0$
and $\phi(S)<S$, there holds $\sum_{i=1}^{\sfM}\pi_i
J{W}^{(i)}JSJ({W}^{(i)})'J<JSJ+\mathbf{11}'$.
Finally, let $Z=S$ and $Y=\tau_\ast  S$.
By Schur complement lemma, we see that~\eqref{eqn:LMI-1} and~\eqref{eqn:LMI-3} hold.
 In addition, the optimization is a
generalized eigenvalue problem, which is
quasiconvex~\cite{boyd1994lmi}.

\textit{Sufficiency:}
For any given $\tau_\ast\leq\tau_\dagger$,
there always exist $Y$ and $Z$ such
that~\eqref{eqn:LMI-1},~\eqref{eqn:LMI-2}~and~\eqref{eqn:LMI-3}
hold. According to Schur complement lemma,~\eqref{eqn:LMI-1}
is equivalent to
\begin{equation*}
JZJ+\mathbf{11}'-\sum_{i=1}^{\sfM}\pi_i
(JZ- JL^{(i)}JY)Z^{-1}(JZ- JL^{(i)}JY)^*>0,
\end{equation*}
which gives
\begin{align}\label{eqn:inequailty-from-LMI}
JZJ+\mathbf{11}'&>JZJ+\sum_{i=1}^{\sfM}\pi_i
\Bigl[
 JL^{(i)}JYZ^{-1}YJ(L^{(i)})'J
-JYJ(L^{(i)})'J-JL^{(i)}JYJ\Bigl]\notag\\
&\geq JZJ+\sum_{i=1}^{\sfM}\pi_i
\Bigl[
 \tau_\ast JL^{(i)}JYJ(L^{(i)})'J
-JYJ(L^{(i)})'J-JL^{(i)}JYJ\Bigl]\notag\\
&\geq JZJ-{\tau_\ast}^{-1}JYJ+{\tau_\ast}^{-1}\phi(Y),
\end{align}
where the second inequality holds by substituting $Z^{-1}$ with $\tau_\ast Y^{-1}$ in accordance with~\eqref{eqn:LMI-3}.
Therefore, it leads to $JYJ+{\tau_\ast}\mathbf{11}'>\phi(Y)$.
Letting $S=JYJ+{\tau_\ast}\mathbf{11}'$, we have
$$\phi(Y)=\sum_{i=1}^{\sfM}\pi_i
JW^{(i)}J(JYJ+{\tau_\ast}\mathbf{11}')J(W^{(i)})'J=\phi(S)$$
and $S>\phi(S)$. In addition,
the positive definiteness of $S$ can be seen from the following lemma.
\begin{lemma}\label{lemma:JXJ-large-X}
There holds $JMJ+\epsilon \mathbf{11}'>0$ for all $M> 0$ and
$\epsilon>0$, where
$J$ is defined in~\eqref{def:J-matrix}.
\end{lemma}
\textit{Proof.}
Choose for $(v_1,\ldots, v_{\sfN})$ an orthonormal basis with $v_{1}=\frac{1}{\sfN}\mathbf{1}$. For any
nonzero vector $x=\sum_{i=1}^{\sfN}a_iv_i$,
\begin{equation*}
x'(JMJ+\epsilon \mathbf{11}')x=
\Big(\sum_{i=2}^{\sfN}a_iv_i\Big)'M\Big(\sum_{i=2}^{\sfN}a_iv_i\Big)
+\epsilon a_1^2.
\end{equation*}
Since $a_1,\ldots, a_{\sfN}$ are not all $0$
and $M>0$, we have $x'(JMJ+\epsilon \mathbf{11}')x>0$.
\hfill $\square$\\
By Proposition~\ref{thm:iff-mean-square-consensus},
Algorithm \eqref{eqn:evolution-xt} achieves mean-square consensus,
which completes the proof.
\hfill $\square$

\subsection{Almost Sure Consensus/Divergence}

In this part, we focus on the impact of sampling intervals on almost sure consensus and almost sure divergence of Algorithm \eqref{eqn:evolution-xt}. The following theorem gives the relationships between $\tau_\ast$ and almost sure consensus/divergence: almost sure
divergence is achieved when $\tau_\ast$ exceeds an upper bound and almost sure consensus is guaranteed when $\tau_\ast$ is sufficiently small. Also note these two boundaries are not equal in general.
\begin{theorem}\label{thm:thm3-as-consensus-iid}
Let Assumptions~(A1),~(A2), and (A3) hold.
\begin{itemize}
\item[(i)] If $\tau_\ast\leq\tau_\dagger$ with $\tau_\ast$
given in Theorem~\ref{prop:propostion1},
 Algorithm \eqref{eqn:evolution-xt} achieves almost sure consensus.

\item[(ii)] If $\tau_\ast>\tau_\natural$, where $\tau_\natural\in\bb{R}_+$ is given by
$$\tau_\natural:=\inf\left\{\tau:\log\frac{2\sfN(\tau-1)}{\sfN-1}>
\frac{(1-q)\log(2\sfN)}{q_\ast q},\mathpzc{s}(\tau)\geq 0
\right\}$$
with $q_\ast:=\min\{(1-q)^{|\mathscr{N}_i|+|\mathscr{N}_j|}:(j,i)\in\mathrm{E}\}$
and
$\mathpzc{s}(\tau):=\min\Big\{\lambda_{\min}
\big(\tau(L^{(i)})'J{L^{(i)}}- J{L^{(i)}}- (L^{(i)})'J\big): {L^{(i)}}\in\mathscr{L}\Big\}$,
Algorithm \eqref{eqn:evolution-xt} diverges almost surely for any initial state
$x(t_0)\in \mathbb{R}^{\mathsf{N}}$ except $x(t_0)\perp \mathbf{1}$.
\end{itemize}
\end{theorem}
\textit{Proof.}
We start by presenting supporting lemmas.
\begin{lemma}[Lemma (5.6.10) in~\cite{horn2012matrix}]
\label{lemma:epsilon-matrix-norm}
Let $A\in\mathbb{C}^{n\times n}$ and $\epsilon>0$ be given.
There is a matrix norm $\|\cdot\|_\dagger$ such that
$\rho(A)\leq\|A\|_\dagger\leq \rho(A)+\epsilon.$
\end{lemma}

\begin{lemma}[Borel-Cantelli Lemma]\label{lemma:borel-cantelli-lemma}
Let $(\mathscr{S}, \mathcal{S}, \mu)$ be a probability space. Assume that events
$\mathscr{A}_i\in \mathcal{S}$ for all $i\in{\mathbb{N}}$.
If $\sum\limits_{i=0}^{\infty}\mu(\mathscr{A}_i)<\infty$, then $\mu\left
(\mathscr{A}_i{~i.o.}\right)=0$, where ``$\mathscr{A}_i~{i.o.}$" means $\mathscr{A}_i$
occurs infinitely often.
In addition, assuming that events $\mathscr{A}_i$, $i\in\mathbb{N}$, are independent, then
$\sum\limits_{i=0}^{\infty}\mu(\mathscr{A}_i)=\infty$ implies
 $\mu\left(\mathscr{A}_i~{i.o.}\right)=1$.
\end{lemma}

\noindent \textit{Proof of (i):}
Note that
$$\mathbb{E}[\|d(k)\|^2]=\Tr\left(\mathbb{E}[d(k)d(k)^*]\right)\leq
\sfN^{1/2}\big\|\vec\left(\mathbb{E}[d(k)d(k)^*]\right)\big\|.$$
The inequality results from the fact that, for any $X:=[x_{ij}]\in\mathbb{S}_+^n$,
$\|\vec(X)\|^2=\sum_{i=1}^n\sum_{j=1}^nx_{ij}^2\geq \sum_{i=1}^{n}x_{ii}^2
\geq \frac{1}{n}(\sum_{i=1}^nx_{ii})^2=\frac{1}{n}(\mathrm{Tr}(X))^2$.
If $\tau_\ast<\tau_\dagger$ or equivalently
$\rho\Big(\mathbb{E}[W(0)\otimes W(0)](J\otimes J)\Big)<1$ by
Theorem~\ref{prop:propostion1},
there exists
a matrix norm $\|\cdot\|_\dagger$ such that
$\bigl\|\mathbb{E}[W(0)\otimes W(0)] (J\otimes J)\bigl\|_\dagger<
\lambda<1$ by Lemma~\ref{lemma:epsilon-matrix-norm}. Moreover, by
the equivalence of norms on a finite-dimensional vector
space, for the two norms
$\|\cdot\|$ and $\|\cdot\|_\dagger$, there exists a real number
$c\in\bb{R}_+$ implying
$$\|X\|\leq c\|X\|_\dagger$$
for all $X\in\bb{R}^{n\times n}$.
From the forgoing observations,~\eqref{eqn:iteration-dk} and
the submultiplicativity of a matrix norm,
\begin{align*}
&\mathbb{E}[\|d(k)\|^2]\\
\leq&\sfN^{1/2}\left\|( J\otimes J)\Big(\mathbb{E}[W(0)\otimes W(0)]( J\otimes J)\Big)^k
\mathrm{vec}\left(x(t_0)x(t_0)^*\right)\right\|\\
\leq&\sfN^{1/2}c\left\|\Big(\mathbb{E}[W(0)\otimes W(0)]
( J\otimes J)\Big)^k\right\|_\dagger
\big\|\mathrm{vec}\left(x(t_0)x(t_0)^*\right)
\big\|\\
\leq&\sfN^{1/2}c\,
\Bigl\|\mathbb{E}[W(0)\otimes W(0)]
( J\otimes J)\Bigl\|_\dagger^k\big\|\mathrm{vec}\left(x(t_0)x(t_0)^*\right)
\big\|\\
<&c\lambda^k\sfN^{1/2}\big\|\mathrm{vec}\left(x(t_0)x(t_0)^*\right)\big\|.
\end{align*}
Therefore,
\begin{equation}\label{eqn:stochastica-stability}
\sum_{k=0}^\infty\mathbb{E}[\|d(k)\|^2]< c(1-\lambda)^{-1}
\sfN^{1/2}\big\|\mathrm{vec}\left(x(t_0)x(t_0)^*\right)
\big\|
<\infty,
\end{equation}
together with Markov's inequality resulting in that,
$$\sum_{k=0}^\infty\bb{P}\big(\|d(k)\|>\delta\big)\leq(1/\delta^2)\sum_{k=0
}^\infty\mathbb{E}[\|d(k)\|^2]
<\infty$$ holds for any $\delta>0$.
According to Lemma~\ref{lemma:borel-cantelli-lemma},
$\lim_{k\to\infty}\|d(k)\|=0$ almost surely for any initial
 state $x(t_0)\in \mathbb{R}^{\mathsf{N}}$.
Then, $\mathbb{P}\left(\lim_{k\rightarrow \infty}\mathfrak{X}(k)=0\right)=1$
follows from~\eqref{eqn:dt-lower-bound}~and~\eqref{eqn:dt-upper-bound}.

\noindent \textit{Proof of (ii):} The rest of the proof consists of three steps. In the first two steps, we construct a sequence of i.i.d. random variables and give a lower bound of
the averaged rate of divergence for this sequence.
In the third step, the strong law of large numbers is
applied to deduce the divergence result.
\\
\noindent \textbf{Step 1}.
First of all, observe that for all $k\in\mathbb{N}$ and $\omega\in\mathscr{G}^{\bb{N}}$
\begin{align*}
\|d(k+1,\omega)\|^2=&
d(k,\omega)^*W(k,\omega)'JJW(k,\omega)d(k,\omega)\\
\geq&{\min_{\|v\|=1,\atop v\perp\mathbf{1}}}
\Big\|v^*W(k,\omega)'JW(k,\omega) v\Big\| \,
\|d(k,\omega)\|^2,
\end{align*}
where the inequality holds because $d(k,\omega)\perp \mathbf{1}$.
If $\lambda_{\min}\Big(W(k,\omega)'JW(k,\omega)+\frac{1}{\sfN}\mathbf{11}'\Big)\geq 1$
for any $k\in\mathbb{N}$ and $\omega\in\mathscr{G}^{\bb{N}}$, then ${\min_{\|v\|=1,\atop v\perp\mathbf{1}}}
\Big\|v^*W(k,\omega)'JW(k,\omega) v\Big\| \geq 1$,
which together with~\eqref{eqn:dt-lower-bound}~and~\eqref{eqn:dt-upper-bound} implies that
\begin{equation}
\bb{P}\left({\mathfrak{X}^2(k)}\geq
\frac{{\mathfrak{X}^2(k-1)}}{2\sfN}\right)=1
\end{equation}
holds for all $k\in\bb{N}$. {Therefore, $\mathfrak{X}(k)>0$
 for all $k\in\mathbb{N}$ provided that $\mathfrak{X}(0)>0$. The following random variables are well defined:
$$\xi(k):=\frac{\mathfrak{X}^2(k+1)}{\mathfrak{X}^2(k)},
~~~k\in\mathbb{N}.$$
}

One condition guaranteeing $\lambda_{\min}\Big(W(k,\omega)'JW(k,\omega)+\frac{1}{\sfN}\mathbf{11}'\Big)\geq 1$
is established as follows. Note that for any ${L^{(i)}}\in\mathscr{L}$,
\begin{align}
&\lambda_{\min}\Big(W(k,\omega)'JW(k,\omega)+\frac{1}{\sfN}\mathbf{11}'\Big)\\
=&
\lambda_{\min}\Big(\tau^2(L^{(i)})'J{L^{(i)}}-\tau J{L^{(i)}}-\tau
(L^{(i)})'J+I\Big)\notag\\
=&
\tau\lambda_{\min}\Big(\tau(L^{(i)})'J{L^{(i)}}- J{L^{(i)}}-
(L^{(i)})'J\Big)+1.
\end{align}
Introduce \begin{equation}\label{eqn:tau-sharp}
\tau_\sharp=\inf\left\{\tau:\lambda_{\min}\Big
(\tau(L^{(i)})'J{L^{(i)}}- J {L^{(i)}}- (L^{(i)})'J\Big)\geq 0,~\forall {L^{(i)}}\in\mathscr{L}\right\}.
\end{equation}
A basic but vital observation is that $\tau_\sharp<\infty$, which
makes $\tau_\sharp$ well defined. To see this, choose a positive number
$\tau_i$ associated with any given ${L^{(i)}}\in\mathscr{L}$. If $\lambda_{\min}\Big(\tau(L^{(i)})'J{L^{(i)}}- J {L^{(i)}}- (L^{(i)})'J\Big)\geq 0$ for any $L^{(i)}$, we are done.
Otherwise let $v\in\bb{C}^{\sfN}$
with $\|v\|=1$ be any vector such that
\begin{equation}\label{eqn:quadradic-form}
v^*\Big(\tau_i(L^{(i)})'J{L^{(i)}}- J{L^{(i)}}- (L^{(i)})'J\Big)v<0.
\end{equation}
Using the property $\ker((L^{(i)})'J{L^{(i)}})=\ker(J{L^{(i)}})$, we deduce from~\eqref{eqn:quadradic-form}
$v\not\in\ker((L^{(i)})'J{L^{(i)}})$ and $v^*(L^{(i)})'J{L^{(i)}}v>0$.
Let $\tau_i$ take a new value satisfying
$$\tau_i> \frac{\max_{\|v\|=1}v^*(J{L^{(i)}}+(L^{(i)})'J)v}
{\min_{\|v\|=1,v\not\in\ker(J{L^{(i)}})} v^*(L^{(i)})'J{L^{(i)}}v}\geq
\sup_{\|v\|=1,v\not\in\ker(J{L^{(i)}})}\frac{v^*( J{L^{(i)}}+ (L^{(i)})'J)v}
{v^*(L^{(i)})'J{L^{(i)}}v}.$$
Then, $\lambda_{\min}\Big(\tau_i(L^{(i)})'J{L^{(i)}}- J{L^{(i)}}- (L^{(i)})'J\Big)\geq 0$. Finally, letting $\tau_0=\max_{1\leq i\leq n}\tau_i$, we have
$\tau_\sharp\leq \tau_0<\infty$.
According to Weyl Theorem (Theorem 4.3.1 in~\cite{horn2012matrix}),
$\lambda_{\min}\Big(\tau(L^{(i)})'J{L^{(i)}}- J{L^{(i)}}-
(L^{(i)})'J\Big)\geq0$
whenever $\tau>\tau_\sharp$ for each ${L^{(i)}}\in\mathscr{L}$. Recalling that
$L(k,\omega)\in\mathscr{L}$, we see that
$\tau>\tau_\sharp$ guarantees $\lambda_{\min}\Big(W(k,\omega)'JW(k,\omega)+\frac{1}{\sfN}\mathbf{11}'\Big)\geq 1$
for all $k\in\mathbb{N}$ and $\omega\in\mathscr{G}^{\bb{N}}$ .

\noindent \textbf{Step 2}. First, we propose the following claim.

\textit{Claim.}~There always exist
two (random) nodes $i,j\in\mathrm{V}$ at each time $k$ such that $(j,i)\in\mathrm{E}$
and $|x_i(t_k)-x_j(t_k)|\geq \frac{1}{\sfN-1}\mathfrak{X}(k)$.

To prove the claim, fix any time instance $k$. Without loss of generality, index all the nodes in the graph such that
$x_{\min}(t_k)=x_{i_1}(t_k)\leq x_{i_2}(t_k)\leq \cdots\leq
x_{i_\sfN}(t_k)=x_{\max}(t_k)$. Then, there at least exists
a node $i_n\in\{i_2,\ldots,i_{\sfN}\}$ satisfying $x_{i_n}(t_k)-x_{i_{n-1}}(t_k)\geq \frac{1}{\sfN-1}\mathfrak{X}(k)$; otherwise $x_{i_\sfN}(t_k)-x_{i_1}(t_k)=\sum_{l=2}^{\sfN}(x_{i_l}
(t_k)-x_{i_{l-1}}(t_k))<\mathfrak{X}(k)$, reaching a contradiction.
If $|x_i(t_k)-x_j(t_k)|< \frac{1}{\sfN-1}\mathfrak{X}(k)$ for all $(j,i)\in\mathrm{E}$, then neither $(i,j)\not\in\mathrm{E}$ nor
$(j,i)\not\in\mathrm{E}$ for any
$i=i_1,\ldots,i_{n-1}$ and $j=i_n,\ldots,i_{\sfN}$, since
$x_{i_n}(t_k)-x_{i_{n-1}}(t_k)\geq \frac{1}{\sfN-1}\mathfrak{X}(k)$,
contradicting with the hypothesis that $\mathrm{G}$ has a directed spanning tree.

In view of this claim, for each $\omega\in\mathscr{G}^{\bb{N}}$,
we choose two nodes $i_k(\omega),j_k(\omega)\in\mathrm{V}$ at time $k$
such that $(j_k(\omega),i_k(\omega))\in\mathrm{E}$ and
 $|x_{i_k(\omega)}(t_k)-x_{j_k(\omega)}(t_k)|\geq \frac{1}{\sfN-1}\mathfrak{X}(k,\omega)$.
 The dependence of the node selections
 on a specific sample path gives rise to a challenge in
 the subsequent analysis.
To get rid of this, we introduce an additional sequence of random
variables. Let $\{z_k\}_{k\in\bb{N}}$ be a sequence of i.i.d. random variables defined on $\left((0,1)^{\bb{N}}, (\mathcal{B}(0,1))^{\bb{N}},\mathpzc{l}\right)$, where
$\mathcal{B}(0,1)$ denotes the Borel algebra on $(0,1)$,
 with $z_k(\zeta)=\zeta_k$ for
all $\zeta\in(0,1)^{\bb{N}}$
and
each $z_k$ uniformly distributed in $(0,1)$.
Let $z_0,z_1,\ldots$ and $\mathrm{G}_0, \mathrm{G}_1,\ldots$ be independent.
Formally, we are allowed to
define a product probability space $(\mathscr{S},\mathcal{S}, \mu)$
where $\mathscr{S}=\mathscr{G}^{\bb{N}}\times (0,1)^{\bb{N}}$,
$\mathcal{S}$ is the $\sigma$-algebra generated by
$\big\{\mathscr{A}\times\mathscr{B}: \mathscr{A}\in\mathcal{F},
\mathscr{B}\in(\mathcal{B}(0,1))^{\bb{N}}\big\}$, and
$\mu$ is the probability measure satisfying
$\mu(\mathscr{A}\times\mathscr{B})=\bb{P}(\mathscr{A})\mathpzc{l}(\mathscr{B})$.
Define $\mathcal{S}_k=\sigma\left((\mathrm{G_0},z_0),\ldots,
(\mathrm{G}_k,z_k)\right)$.
Introduce a sequence of events
associated with $i_k(\omega)$, $j_k(\omega)$ and
$z_k$:
$$
\mathscr{D}(k)=\Big\{
\cup_{\omega\in\mathscr{G}^{\bb{N}}}(\omega\times\mathscr{B}_k(\omega)):\mathscr{N}_{i_k(\omega)}
(k,\omega)=\{j_k(\omega)\},
\mathscr{N}_{j_k(\omega)}(k,\omega)=\emptyset
\Bigl\}$$
with $\mathscr{B}_k({\omega})=\left\{\zeta\in(0,1)^{\bb{N}}:z_k(\zeta)<
q_\ast/(1-q)^{|\mathscr{N}_{i_k(\omega)}|+
|\mathscr{N}_{j_k(\omega)}|}\right\}$.
Since $i_k(\omega),j_k(\omega)\in\mathcal{F}_{k-1}$,
one can verify $\mathscr{D}(k)\in\mathcal{S}_k$.
If $\tau_\ast>1$, for all $(\omega, \zeta)\in\mathscr{D}(k)$
and $k\in\bb{N}$,
\begin{align}\label{eqn:diverge-factor}
\mathfrak{X}(k+1,\omega)\geq&
|x_{i_k(\omega)}(t_{k+1})-x_{j_k(\omega)}(t_{k+1})|\notag\\
=&(\tau_\ast-1)
|x_{i_k(\omega)}(t_{k})-x_{j_k(\omega)}(t_{k})|\notag\\
\geq &\frac{\tau_\ast-1}{\sfN-1}\mathfrak{X}(k,\omega).
\end{align}
 Direct calculation yields
\begin{equation}
\mu((\omega, \zeta)\in\mathscr{D}(k))= \frac{q_\ast q}{1-q}.
\end{equation}

\noindent \textbf{Step 3}.~Now we define random variables
\begin{equation}\label{eqn:def-Mk}
\mathfrak{M}(k)=\left\{
\begin{array}{cl}
\frac{\tau_\ast-1}{\sfN-1}, & \hbox{if~}(\omega, \zeta)\in\mathscr{D}(k),\\
\frac{1}{2\sfN},&\hbox{otherwise;}
\end{array}\right.
\end{equation}
which together with~\eqref{eqn:diverge-factor} leads to
\begin{equation*}
\mu\left(\xi_k=\frac{\mathfrak{X}^2(k+1)}{\mathfrak{X}^2(k)}\geq
\mathfrak{M}^2(k)\right)=1.
\end{equation*}
Therefore,
\begin{equation*}
\mu\left(\prod_{k=0}^{t}\xi_k=
\frac{\mathfrak{X}^2(t+1)}{\mathfrak{X}^2(0)}\geq
\prod_{k=0}^{t}\mathfrak{M}^2(k)\right)=1,
\end{equation*}
which gives
\begin{equation}\label{eqn:Xk-upper-bound}
\mu\left(
\log\mathfrak{X}(t+1)-\log{\mathfrak{X}(0)}\geq
\sum_{k=0}^{t}\log \mathfrak{M}(k)\right)=1.
\end{equation}
Since each node samples the neighbors independently, where
the ``independence" is in both spatial and temporal sense (Assumption (A2)), therefore, for any $k\in\bb{N}$,
$$\mu\left((\omega,\zeta)\in\mathscr{D}(k)\mid
\mathcal{S}_{k-1}\right)=\frac{p_\ast p}{1-p}=\mu\left((\omega,\zeta)\in\mathscr{D}(k)\right),$$
indicating that
$\mathfrak{M}(k)$'s are independent random variables for
$\mathscr{D}(0),\ldots,\mathscr{D}(k-1)\in\mathcal{S}_{k-1}$.
By induction, we eventually have $\{\mathfrak{M}(k)\}_{k\in\mathbb{N}}$ are i.i.d.
with the mean computed as
\begin{equation}\label{eqn:lower-bound-ElogMk}
\bb{E}[\log \mathfrak{M}(k)]=
\frac{q_\ast q}{1-q}\log\frac{\tau_\ast-1}{\sfN-1}+
(1-\frac{q_\ast q}{1-q})\log\frac{1}{2\sfN}:=\mathpzc{m}(\tau_\ast).
\end{equation}
Additionally, since $\mathfrak{M}(k)$'s have uniformly bounded covariances, Kolmogorov's strong law of large numbers~\cite{feller1968introduction}
shows that
\begin{equation}\label{eqn:slln-iid}
\mu\left(\lim_{t\to\infty}\frac{1}{t}
\sum_{k=0}^t\log\mathfrak{M}(k)=\mathpzc{m}(\tau_\ast)\right)=1,
\end{equation}
which together with~\eqref{eqn:Xk-upper-bound} implies that, when
$\mathpzc{m}(\tau_\ast)>0$, $\mathbb{P}\left(\liminf_{k\to\infty}\mathfrak{X}(k)=\infty\right)=1.$
Notice that $\mathpzc{m}(\tau_\ast)$ is increasing in $\tau_\ast$.
Defining
$\tau_\flat=\inf\left\{\tau:\mathpzc{m}(\tau_\ast)>0
\right\}$
and choosing $\tau_\ast> \tau_\sharp\vee \tau_\flat:=\tau_{\natural}$, the conclusion follows.
\hfill $\square$

\section{Markovian Random Networks}\label{sec:markovian-sampling}
In this section, we continue to
investigate the sampled-data consensus when each node samples the neighbors following a Markov chain. The following assumption is imposed.
\begin{enumerate}
\item[(A4)]\label{asmpt:assumpt-MC}
 \emph{Independently among
$(j,i)\in \mathrm{E}$, the random variables
$\mathbf{1}_{\{(j,i)\in \mathrm{E}_k\}}$, $k=0,1,\ldots,$
are a binary Markov chain with the failure rate
$\mathbb{P}\left(  \mathbf{1}_{\{(j,i)\in \mathrm{E}_{k+1}\}}=0
 \mid\mathbf{1}_{\{(j,i)\in \mathrm{E}_k\}}=1\right):=p$
and the recovery rate
$\mathbb{P}\left(  \mathbf{1}_{\{(j,i)\in \mathrm{E}_{k+1}\}}=1
 \mid\mathbf{1}_{\{(j,i)\in \mathrm{E}_k\}}=0\right):=q$ positive and
 strictly less than one.}
\end{enumerate}
Note that the techniques developed
in this section also apply when $p(i)$ and $q(i)$ vary depending on the
node index $i$.

Under Assumption~(A4),
$\{{L}(k)\}_{k\in\bb{N}}$ are a sequence of
random variables taking values from $\mathscr{L}$, governed by a finite-state
time-homogeneous Markov chain.
The transition probability of $\{{L}(k)\}_{k\in\mathbb{N}}$ is
induced from the transition of edges between the ``on'' state and the ``off'' state, which is
\begin{equation}\label{def:Pi-matrix}
\mathbb{P}\Bigl(L(k)={L^{(j)}}\mid L(k-1)={L^{(i)}}\Bigl)=
p^{s_1}
(1-p)^{s_2} q^{s_3}
(1-q)^{s_4}
:=\pi_{ij}.
\end{equation}
where
$s_1=\sum\limits_{(n,m)\in\mathrm{E}}\mathbf{1}_{\{l_{mn}^{(i)}\not=0,
l_{mn}^{(j)}=0\}}$, $s_2=\sum\limits_{(n,m)\in\mathrm{E}}\mathbf{1}_{\{l_{mn}^{(i)}\not=0,
l_{mn}^{(j)}\not=0\}}$,
$s_3=\sum\limits_{(n,m)\in\mathrm{E}}\mathbf{1}_{\{l_{mn}^{(i)}=0,
l_{mn}^{(j)}\not=0\}}$, and
$s_4=\sum\limits_{(n,m)\in\mathrm{E}}\mathbf{1}_{\{l_{mn}^{(i)}=0,
l_{mn}^{(j)}=0\}}$.

For convenience, we denote $\Pi:=[\pi_{ij}]$ as the \textit{transition probability matrix} of $L(k)$.
Again, $W(k)$ inherits the same distribution from ${L}(k)$.
The positiveness of the recovery and failure rates in Assumption (A4)
makes ${L}(k)$ an ergodic Markov chain and
$\Pi$ a positive matrix.

\subsection{Conjunction of Various Consensus Metrics}
In this part, we show that
 an analog of Theorem~\ref{thm:thm1} holds over a Markovian random network.
From the probabilistic point of view,
the difference between independent model and
Markovian model can be interpreted using a finite permutation argument
as follows.
Let $\mathpzc{q}$ be a finite permutation from $\mathbb{N}$ onto
$\mathbb{N}$ such that $\mathpzc{q}(i)\not=i$ for finitely many $i$.
For any given $\omega\in\mathscr{G}^{\mathbb{N}}$,
we define a finite permutation
as $(\mathpzc{q} \omega)_i=\omega_{\mathpzc{q}(i)}$ for all $i\in\mathbb{N}$. In the i.i.d. model, the probability measure is
invariant with respect to a finite permutation of the sample path, i.e.,
$\mathbb{P}(\omega\in\mathcal{F})=\mathbb{P}
(\omega\in\mathpzc{q}\mathcal{F})$; while in the
Markovian model this property is absent because of the Markov property.
Nevertheless, if $\tau_k=\tau_\ast\in\big(0,(\mathsf{N}-1)^{-1}\big)$
for all $k$, the difference does not play any key role in whether or not
$\mathfrak{X}(k)$ converges in expectation for Algorithm~\eqref{eqn:evolution-xt}. Moreover,
Lemma~\ref{lemma:equ-consensus-notaions} guarantees mean-square consensus, and almost sure consensus regardless of the random network model.
\begin{theorem}\label{thm:thm4}
Let Assumptions~(A1),~(A3), and~(A4) hold.
Then expectation consensus,
mean-square consensus, and almost sure consensus
are achieved for Algorithm \eqref{eqn:evolution-xt} if
$\tau_\ast\in\bigl(0,(\mathsf{N}-1)^{-1}\bigl)$.
\end{theorem}
\textit{Proof.}
The proof is similar to that of Theorem~\ref{thm:thm1}.
Here we only provide a sketch.
Fix a directed spanning tree $\mathrm{G}_T$ of
graph $\mathrm{G}$ and a sampling time $t_k$.
We choose
$i_1,\ldots,i_{\mathsf{N}}$ and define
$\mathscr{M}_1,\ldots,\mathscr{M}_{\mathsf{N}}$ in
sequel by the following iterated algorithm:
\begin{inparaenum}[\itshape 1)]
\item Set $i_1$ as the root node of $\mathrm{G}_T$,
$\mathscr{M}_1:=\{i_1\}$ and $l=2$;
\item\label{step:2} Choose a node $i_l\in\mathrm{V}$ such that
 there exists
a node $j\in\mathscr{M}_{l-1}$ satisfying
$(j,i_l)\in\mathrm{G}_T$ and
$i_l\not\in\mathscr{M}_{l-1}$;
\item
Update $\mathscr{M}_l:=
\mathscr{M}_{l-1}\cup\{i_l\}$;
\item If $l\leq\sfN$, set $l= l+1$ and \text{go to}~step~$\mathit{2)}$;
otherwise \text{stop}.
\end{inparaenum}
Consider a sequence of events
$\mathscr{E}_2,\ldots,\mathscr{E}_{\sfN}$ where
$\mathscr{E}_l:=\left\{L(k+l-1)\in\{
L^{(j)}\in\mathscr{L}:l_{i_li_{l-1}}^{(j)}
\not=0\}
\right\}$ for $l=2,3,\dots, \sfN$.
If $\mathscr{E}_2,\ldots,\mathscr{E}_{\sfN}$
sequentially occur, similar to the proof of Theorem~\ref{thm:thm1}, we see
that \begin{equation}
\mathfrak{X}({k+\mathsf{N}-1})
\leq\left(1-\frac{1}{2}\eta^{\mathsf{N}-1}\right)\mathfrak{X}(k)
\end{equation}
where $\eta:=( \tau_\ast)\wedge(1-(\sfN-1) \tau_\ast)>0$.
Then, we estimate the probability of the sequential occurrence of
 $\mathscr{E}_2,\ldots,\mathscr{E}_3$ by
\begin{align*}
\mathbb{P}\left({1}_{\cap_{i=2}^{\mathsf{N}}
\mathscr{E}_i}=1\mid L(k-1)\in\mathscr{L}\right)=&
\mathbb{P}\left({1}_{\mathscr{E}_\sfN}=1\mid
{1}_{\mathscr{E}_{\sfN}-1}=1\right)
\cdots\mathbb{P}\left({1}_{\mathscr{E}_2}=1\mid
L(k-1)\in\mathscr{L}\right)\\
\geq&
\pi^{\mathsf{N}-1},
\end{align*}
where $\pi:=\min_{1\leq i,j\leq \sfM}\pi_{ij}>0.$ Therefore,
\begin{align*}
&\mathbb{E}[\mathfrak{X}({k+\mathsf{N}-1})]\\
=&
\sum_{\gamma=0,1}\mathbb{E}\biggl[
\mathbb{P}\bigl(\mathbf{1}_{\cap_{i=2}^{\mathsf{N}}
\mathscr{E}_i}=\gamma\mid L(k-1)\bigl)
\mathbb{E}\Bigl[\mathfrak{X}(k+\sfN-1)\mid
\mathbf{1}_{\cap_{i=2}^{\mathsf{N}}
\mathscr{E}_i}=\gamma,L(k-1)\Bigl]
\biggl]\\
\leq&
(1-\pi^{\mathsf{N}-1})\mathbb{E}\bigl[\mathbb{E}[\mathfrak{X}(k)\mid L(k-1)]\bigl]
+\pi^{\mathsf{N}-1}\left(1-\frac{1}{2}\eta^{\mathsf{N}
-1}\right)\mathbb{E}\bigl[\mathbb{E}[\mathfrak{X}(k)\mid L(k-1)]\bigl]\\
=&\left(1-\frac{1}{2}(\pi \eta)^{\mathsf{N}-1}\right)
\mathbb{E}[\mathfrak{X}(k)],
\end{align*}
which implies $\lim_{k\to\infty}\mathbb{E}[\mathfrak{X}(k)]=0$ and
therefore consensus in expectation is achieved. Finally, the conclusion follows from Lemma \ref{lemma:equ-consensus-notaions}.
\hfill $\square$
\begin{remark}
The assumption of a uniform inter-sampling interval $\tau_k$
simplifies the notations used in Theorems~\ref{thm:thm1} and~\ref{thm:thm4}.
It should be emphasized that the techniques used in the proof of
Theorems~\ref{thm:thm1} and~\ref{thm:thm4} also apply to the non-uniform inter-sampling
interval case. To make the conclusion hold, we require $\lim_{k\to\infty} \log\big(\mathbb{E}[\mathfrak{X}\big(k(\sfN-1)\big)]
/\mathbb{E}[\mathfrak{X}(0)]\big)
=-\infty$, which can be guaranteed by $\sum_{k=0}^\infty
\prod_{j=0}^{\sfN-2} \eta_{k+j}=\infty$ with
$\eta_k=( \tau_k)\wedge(1-(\sfN-1) \tau_k)$ for
$k\in\bb{N}$. This is seen from~\eqref{eqn:coverge-rate} and the fact that, for
a sequence $\{a_k\}_{k\in\bb{N}}$ with
$a_k\in[0,1)$, $\sum_{k=1}^\infty a_k=\infty$
if and only if $\prod_{k=1}^\infty(1-a_k)=0$~\cite{rudin1987real}.
\end{remark}

\subsection{The Mean-square Consensus Threshold}
Now, we are interested
in establishing a necessary and sufficient condition on $\tau_*$ for mean-square consensus of Algorithm \eqref{eqn:evolution-xt}. We first present an implicit condition
in terms of the spectral radius of a certain matrix. Then,
this stability condition is
translated to a threshold on $\tau_\ast$. The analysis in this section is based on the
techniques using in the proof of Proposition~\ref{thm:iff-mean-square-consensus} as well as
the tools from the theory of Markov jump linear systems.

\begin{proposition}\label{thm:iff-mean-square-consensus-markov}
Let Assumptions~(A1), ~(A3),~and~(A4) hold and,
for each $(j,i)\in \mathrm{E}$,
$\mathbf{1}_{\{(j,i)\in \mathrm{E}_0\}}$ starts at any
initial distribution. Then
the following statements are equivalent:
\begin{itemize}
\item[(i)] Algorithm \eqref{eqn:evolution-xt} achieves  mean-square consensus;

\item[(ii)] There holds
$\rho(\Gamma\Theta)<1$, where
\begin{equation}\label{eqn:def-Theta}
\Gamma:=\diag\Big(W^{(1)}\otimes W^{(1)},\ldots,
W^{(\sfM)}\otimes W^{(\sfM)}\Big)
\end{equation}
and \begin{equation}\label{eqn:def-Pi}
\Theta:=\Pi'\otimes(J\otimes J)
\end{equation}
with $J$ defined in~\eqref{def:J-matrix} and $\Pi$ defined in~\eqref{def:Pi-matrix};

\item[(iii)] There exist matrices $S_1,\ldots,S_{\sfM}>0$ such that
\begin{equation}
\varphi_j(S):=\sum_{i=1}^{\sfM}\pi_{ij}
JW^{(j)}JS_iJ(W^{(j)})'J<S_j
\end{equation}
holds for all $1\leq j\leq \sfM$, where $S:=(S_1,\ldots,S_{\sfM})$.
\end{itemize}
\end{proposition}
\textit{Proof.}
Recall $d(k)$ from~\eqref{eqn:def-dk}.
Obviously,~\eqref{eqn:dt-lower-bound} to~\eqref{eqn:evolution-d_kt}
still hold. In what follows, we consider a linear space over the complex field $\mathbb{C}$: $\mathbb{H}^{\sfM}:=\Big\{(M_1,\ldots,M_\sfM):M_i\in\mathbb{C}^{\sfN\times \sfN},i=1,\ldots,\sfM\Big\}$ and a convex cone in $\mathbb{H}^{\sfM}$:
$\mathbb{H}_+^{\sfM}:=\Big\{(G_1,\ldots,G_\sfM):G_i\in\mathbb{S}_{\sfN}^+,
i=1,\ldots,\sfM\Big\}$.
Define
$$H(k):=\Big(d(k)d(k)^*{1}_{\{L(k)=L^{(1)}\}},\ldots,
d(k)d(k)^*{1}_{\{L(k)=L^{(\sfM)}\}}\Big)\in\mathbb{H}_+^{\sfM}.$$
Since
$$\mathbb{E}[d(k)d(k)^*]=\mathbb{E}[H(k)]
\bigl[\underbrace{I_{\sfN},\ldots,I_{
\sfN}}_{\sfM~\hbox{times}}\big]',$$
 it follows from \eqref{eqn:dt-lower-bound}~and~\eqref{eqn:dt-upper-bound}
 that $\lim_{k\to\infty}\mathbb{E}[\mathfrak{X}^2(k)]=0$ is equivalent to $\lim_{k\to\infty}\mathbb{E}[H(k)]=0$.
Taking vectorization on both side of $\mathbb{E}[H(k)]$ gives
\begin{align*}
\mathrm{vec}\big(\mathbb{E}[H(k)]\big)
=&\left[\begin{array}{ccc}\pi_{11}&\cdots &\pi_{\sfM 1}\\
\vdots & \ddots & \vdots\\
\pi_{1\sfM} & \cdots & \pi_{\sfM \sfM}\end{array}
\right]\otimes
\left[\begin{array}{ccc}(JW^{(1)})\otimes(JW^{(1)})&  & \\
 & \ddots & \\
 & & (JW^{(\sfM)})\otimes(JW^{(\sfM)})\end{array}
\right]\\
&\hspace{49mm}\cdot\left[
\begin{array}{c}
\mathrm{vec}(\bb{E}[d(k-1)d(k-1)^*{1}_{\{L(k-1)=L^{(1)}\}}])\\
\vdots\\
\mathrm{vec}(\bb{E}[d(k-1)d(k-1)^*{1}_{\{L(k-1)=L^{(\sfM)}\}}])
\end{array}
\right]\\
=& \Big((\Pi'\otimes I_{{\sfN}^2})\big(I_{\sfM}\otimes(J\otimes J)\big)\Gamma\Big)
\mathrm{vec}\big(\mathbb{E}[H(k-1)]\big)\notag\\
=&
\Big((\Pi'\otimes I_{{\sfN}^2})\big(I_{\sfM}\otimes(J\otimes J)\big)\Gamma\Big)^k \mathrm{vec}\left(\mathbb{E}[H(0)]\right)\notag\\
=&\Big(\big(I_{\sfM}\otimes(J\otimes J)\big)(\Pi'\otimes I_{{\sfN}^2})\Gamma\Big)^k\big(I_{\sfM}\otimes(J\otimes J)\big)
\,\mathrm{vec}\Big(\mathbb{E}\left[\mathpzc{h}\big(x(t_0),L(0)\big)\right]\Big)\notag\\
=&\big(I_{\sfM}\otimes(J\otimes J)\big)
\Big((\Pi'\otimes I_{{\sfN}^2})\Gamma\big(I_{\sfM}\otimes(J\otimes J)\big)\Big)^k
\mathrm{vec}\Big(\mathbb{E}\left[\mathpzc{h}\big(x(t_0),L(0)\big)\right]\Big),
\end{align*}
where $\mathpzc{h}\big(x(t_0),L(0)\big):=\Big(x(t_0)x(t_0)^*{1}_{\{L(0)=L^{(1)}\}},\ldots,
x(t_0)x(t_0)^*{1}_{\{L(0)=L^{(\sfM)}\}}\Big)\in\mathbb{H}_+^\sfM.$
The fourth equality holds because
$({A} \otimes {B})({C} \otimes {D}) = ({AC}) \otimes ({BD})$
for matrices $A,~B,~C$ and $D$ of compatible dimensions.
In addition,
 $\rho\Big((P'\otimes I_{{\sfN}^2})\Gamma\big(I_{\sfM}\otimes(J\otimes J)\big)\Big)=\rho(\Gamma\Theta)$.

It follows from~Lemma~\ref{lemma:matrix-depomositation} that
for any $H\in\mathbb{H}^{\sfM}$, there exist
$H_1,\ldots,H_4\in\mathbb{H}_+^{\sfM}$ such that
$H=(H_1-H_2)+(H_3-H_4)\mathbf{i}$.
Moreover, for each $H_i=(G_{1}^{(i)},\ldots,G_{\sfM}^{(i)})$,
$$H_i=\sum_{m=1}^{\sfM}\sum_{n=1}^{\sfN}
\lambda_{m,n}^{(i)}\mathpzc{h}(u_{m,n}^{(i)},L^{(m)})$$
with  $U_m^{(i)}=:[u_{m,1}^{(i)},\ldots,u_{m,\mathsf{N}}^{(i)}]$
unitary and
$G_{m}^{(i)}=U_{m}^{(i)}\mathrm{diag}\{\lambda_{m,1}^{(i)},\ldots,
\lambda_{m,\mathsf{N}}^{(i)}\}(U_{m}^{(i)})^*$
 for $m=1,\ldots,\sfM$, which means that, for any $H\in\mathbb{H}^{\sfM}$, $\vec(H)$ can be
 expressed as
a linear combination of different initial states.

 The rest of the proof follows from the arguments used in the proof of Theorem~\ref{thm:thm1} and
 the theory of Markov jump linear systems~\cite{costa2006discrete}.
\hfill $\square$

The following theorem holds based on Theorem~\ref{prop:propostion1} and the theory of Markov jump linear systems, so the proof
is omitted.
\begin{theorem}\label{prop:propostion2}
Let Assumptions~(A1),~(A3), and~(A4) hold.
Then Algorithm \eqref{eqn:evolution-xt} achieves mean-square consensus if and only if $\tau_*\leq \tilde \tau_\dagger$,
where
$\tilde \tau_\dagger$ is given by the following quasi-convex optimization problem:
\begin{align*}
\mathrm{\arg\min}_\tau\, & -\tau \notag\\
\mathrm{subject~to}&\left[\begin{array}{cccc}
JZ_jJ+\mathbf{11}'& \sqrt{\pi_{1j}} \left(JZ_1- J{L}^{(j)}J Y_1\right) & \ldots
&\sqrt{\pi_{\sfM j}}(JZ_\sfM- J
{L}^{(j)}J Y_\sfM)\\
* & Z_1 &\ldots& 0\\
\vdots & \vdots & \ddots &\vdots\\
* &* & \ldots &Z_\sfM \end{array}\right]>0,\notag\\
&Y_j,Z_j>0,\notag\\
&Y_j- \tau Z_j\geq 0, \hspace{2cm}j=1,\ldots,\sfM.\notag
\end{align*}

\end{theorem}

\subsection{Almost Sure Consensus/Divergence}
In this part, we explore the almost sure consensus/divergence condition for Algorithm \eqref{eqn:evolution-xt} over Markovian random networks.
The following theorem exhibits a correlation
between $\tau_\ast$ and the asymptotic behavior of every sample path,
that is, a small $\tau_\ast$ guarantees almost sure consensus while
a large $\tau_\ast$ tends to result in almost sure divergence.
In the following theorem, the almost sure divergence analysis is restricted to complete graphs. The assumption of complete graph
simplifies the analysis. However, we believe that
the techniques used in developing almost sure divergence results
in Theorems~\ref{thm:thm3-as-consensus-iid} and~\ref{thm:thm3-as-consensus-markov} can
also
deal with general directed graphs. We plan to remove this
restriction and consider more general graphs in future work.

\begin{theorem}\label{thm:thm3-as-consensus-markov}
Let Assumptions~(A1),~(A3), and~(A4) hold and,
for each $(j,i)\in \mathrm{E}$,
$\mathbf{1}_{\{(j,i)\in \mathrm{E}_0\}}$ starts at any
initial distribution.
\begin{itemize}
\item[(i)] If $\tau_\ast\leq \tilde \tau_\dagger$,
then Algorithm \eqref{eqn:evolution-xt} achieves almost sure consensus.

\item[(ii)] If $\mathrm{G}$ is a complete graph
and
 $\tau_\ast>\tilde \tau_\natural$, where $\tilde \tau_\natural$ is given by
 $$\tilde \tau_\natural:=\inf\left\{\tau:\tilde q_\ast\log\frac{2\sfN(\tau-1)}{\sfN-1}+
\log\frac{1}{2\sfN}>0,{\mathpzc{s}}(\tau)\geq 0
\right\},$$
with $\tilde q_\ast:=(1-p)\wedge q$
and
${\mathpzc{s}}(\tau):=\min\big\{\lambda_{\min}(\tau
(L^{(i)})'J{L^{(i)}}- J{L^{(i)}}- (L^{(i)})'J): {L^{(i)}}\in\mathscr{L}\big\}$,
then Algorithm \eqref{eqn:evolution-xt} diverges almost surely for any initial state
$x(t_0)\in \mathbb{R}^{\mathsf{N}}$ except $x(t_0)\perp \mathbf{1}$.
\end{itemize}
\end{theorem}
\textit{Proof.}
To show (i), note that
$$\mathbb{E}[\|d(k)\|^2]=\Tr\bigg(\mathbb{E}[H(k)]
\bigl[\underbrace{I_{\sfN},\ldots,I_{
\sfN}}_{\sfM \hbox{~times}}\big]'\bigg)\leq
(\sfM\sfN)^{1/2}\big\|\mathrm{vec}\big(\mathbb{E}[H(k)]\big)\big\|.$$
 When
$\rho(\Gamma\Theta)<1$, by using the same argument as in~\eqref{eqn:stochastica-stability},
we know that
$\sum_{k=0}^\infty\mathbb{E}[\|d(k)\|^2]<\infty$
holds for any initial
 state $x(t_0)\in \mathbb{R}^{\mathsf{N}}$ and any initial distribution of
$\mathbf{1}_{\{(j,i)\in \mathrm{E}_0\}}$ for
each $(j,i)\in \mathrm{E}$.
By Markov's inequality and Lemma~\ref{lemma:borel-cantelli-lemma},
$\lim_{k\to\infty}\|d(k)\|=0$ almost surely.

Next, we shall prove (ii).
Similar to the proof
of Theorem~\ref{thm:thm3-as-consensus-iid}, the analysis is divided into three steps.\\
\noindent \textbf{Step 1}.
Suppose $\tau>\tau_\sharp$, where $\tau_\sharp\in\bb{R}_+$ is
defined in~\eqref{eqn:tau-sharp}. Adopting the analysis
used in the proof of Theorem~\ref{thm:thm3-as-consensus-iid},
we define
$$\xi(k):=\frac{\mathfrak{X}^2(k+1)}{\mathfrak{X}^2(k)},
$$
and conclude that
\begin{equation}
\bb{P}\left({\mathfrak{X}^2(k+1)}\geq
\frac{{\mathfrak{X}^2(k)}}{2\sfN}\right)=1
\end{equation}
holds for all $k\in\bb{N}$.

\noindent \textbf{Step 2}.
In the first place, for each $\omega\in\mathscr{G}^{\bb{N}}$, we choose two (random) nodes  $i_k(\omega),j_k(\omega)\in\mathrm{V}$ at time $k$
such that
 $|x_{i_k(\omega)}(t_k)-x_{j_k(\omega)}(t_k)|= \mathfrak{X}(k,\omega)$. Let $\{z_k\}_{k\in\bb{N}}$ be a sequence of i.i.d. random variables defined on $\left((0,1)^{\bb{N}}, (\mathcal{B}(0,1))^{ \bb{N}},\mathpzc{l}\right)$ with
  $z_k(\zeta)=\zeta_k$ for
all $\zeta\in(0,1)^{\bb{N}}$ and
 each $z_k$ uniformly distributed in $(0,1)$, and let $z_0,z_1,\ldots$ and $\mathrm{G}_0, \mathrm{G}_1,\ldots$ be independent.
Formally, we are allowed to
define a product probability space $(\mathscr{S},\mathcal{S}, \mu)$
with
$\mu$ the product probability measure satisfying
$\mu(\mathscr{A}\times\mathscr{B})=\bb{P}(\mathscr{A})\mathpzc{l}(\mathscr{B})$
for any $\mathscr{A}\in\mathcal{F}$ and
$\mathscr{B}\in(\mathcal{B}(0,1))^{\bb{N}}$.
Define $\mathcal{S}_k=\sigma\left((\mathrm{G_0},z_0),\ldots,
(\mathrm{G}_k,z_k)\right)$.
Introduce a sequence of events
associated with $i_k(\omega)$, $j_k(\omega)$ and
$z_k$:
$$
\mathscr{D}(k)=\Big\{
\cup_{ \omega\in\mathscr{G}^{\bb{N}}}(\omega\times\mathscr{B}_k(\omega)):
(j_k(\omega),i_k(\omega))\in\mathrm{E}_k
\Bigl\}
$$
with $\mathscr{B}_k({\omega})$ given by
\begin{align*}
\mathscr{B}_k({\omega})=\left\{\begin{array}{lll}
\left\{\zeta\in(0,1)^{\bb{N}}:z_k(\zeta)<
((1-p)\wedge q)/(1-p)\right\}, & \hbox{~if~}(j_{k}(\omega),i_{k}(\omega))\in\mathrm{E}_{k-1}(\omega);
\vspace{2mm}
&\\
\left\{\zeta\in(0,1)^{\bb{N}}:z_k(\zeta)<((1-p)\wedge q)/q
\right\}, & \hbox{~if~}(j_{k}(\omega),i_{k}(\omega))\not\in\mathrm{E}_{k-1}(\omega).
\end{array}\right.
\end{align*}
We have the following claim due to
a complete underlying graph $\mathrm{G}$.

\textit{Claim.}~Suppose $\tau_\ast>1$. There holds
$\mathfrak{X}(k+1,\omega)\geq (\tau_\ast-1)
\mathfrak{X}(k,\omega)$ for all $(\omega,\zeta)\in\mathscr{D}(k)$
and $k\in\bb{N}$.

\noindent \textbf{Step 3}.
We define random variables
\begin{equation}\label{eqn:def-Mk}
\mathfrak{M}(k)=\left\{
\begin{array}{cl}
\frac{\tau_\ast-1}{\sfN-1}, & \hbox{if~}(\omega,\zeta)\in\mathscr{D}(k);\\
\frac{1}{2\sfN},&\hbox{otherwise.}
\end{array}\right.
\end{equation}
Similar to the proof of Theorem~\ref{thm:thm3-as-consensus-iid},
for any $t\in\bb{N}$,
\begin{equation}\label{eqn:Xk-upper-bound-mc}
\mu\left(
\log\mathfrak{X}(t+1)-\log{\mathfrak{X}(0)}\geq
\sum_{k=0}^{t}\log \mathfrak{M}(k)\right)=1.
\end{equation}
Since each node independently samples among its neighbors, for any $k\in\bb{N}$,
$$\mu\left((\omega,\zeta)\in\mathscr{D}(k)\mid
\mathcal{S}_{k-1}\right)=(1-p)\wedge q=\mu\left((\omega,\zeta)\in\mathscr{D}(k)\right):=\tilde q_\ast,$$
indicating that
$\mathfrak{M}(k)$'s are independent random variables for
$\mathscr{D}(0),\ldots,\mathscr{D}(k-1)\in\mathcal{S}_{k-1}$.
By induction, we eventually have $\{\mathfrak{M}(k)\}_{k\in\mathbb{N}}$ are i.i.d.
with the mean computed as
\begin{equation}
\bb{E}[\log \mathfrak{M}(k)]\geq
\tilde q_\ast\log\frac{\tau_\ast-1}{\sfN-1}+
(1-\tilde q_\ast)\log\frac{1}{2\sfN}:=\mathpzc{\tilde m}(\tau_\ast)
\end{equation}
In addition, since $\mathfrak{M}(k)$'s have uniformly bounded covariances, again by
Kolmogorov strong law of large numbers~\cite{feller1968introduction},
\begin{equation}\label{eqn:slln-iid}
\mu\left(\lim_{t\to\infty}\frac{1}{t}
\sum_{k=0}^t\log\mathfrak{M}(k)=\bb{E}[\log \mathfrak{M}(k)]\right)=1,
\end{equation}
together with~\eqref{eqn:Xk-upper-bound} implying that, when
$\mathpzc{\tilde m}(\tau_\ast)>0$, $\mathbb{P}\left(\liminf_{k\to\infty}\mathfrak{X}(k)=\infty\right)=1.$
Notice that $\mathpzc{\tilde m}(\tau_\ast)$ is increasing in $\tau_\ast$. The proof is completed by
defining
$\tilde \tau_\flat=\inf\left\{\tau:\mathpzc{\tilde m}(\tau_\ast)>0
\right\}$
and choosing $\tau_\ast> \tau_\sharp\vee \tilde \tau_\flat:=\tilde \tau_{\natural}$.
\hfill $\square$

\section{Numerical Examples}\label{sec:example}
In this section, we provide numerical examples to validate the theoretical results. We first illustrate the existence of the threshold on $\tau_*$, which decides the mean-square convergence or divergence (see Theorems~\ref{prop:propostion1} and~\ref{prop:propostion2}). We then discuss and illustrate how this threshold depends on the number of nodes with cyclic underlying graphs, for i.i.d. and Markovian network models, respectively.


\begin{figure}[t!]
\begin{center}
\includegraphics[width=2.5in]{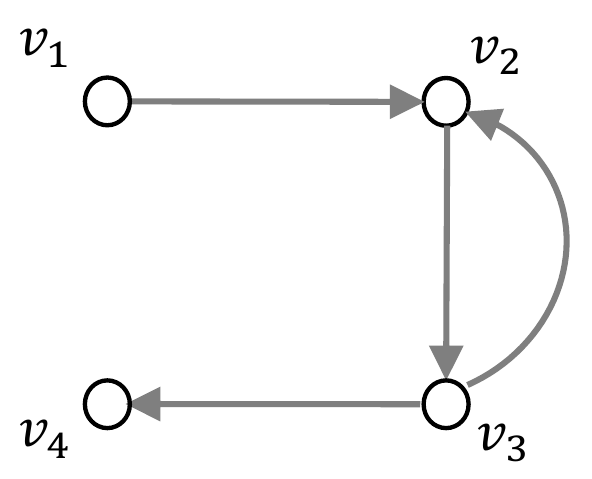}
\caption{A underlying graph~$\mathrm{G}$ consisting of four nodes.}\label{fig:graph-1}
\end{center}
\end{figure}

\begin{figure}[h!]
\begin{center}
\includegraphics[width=5.75in]{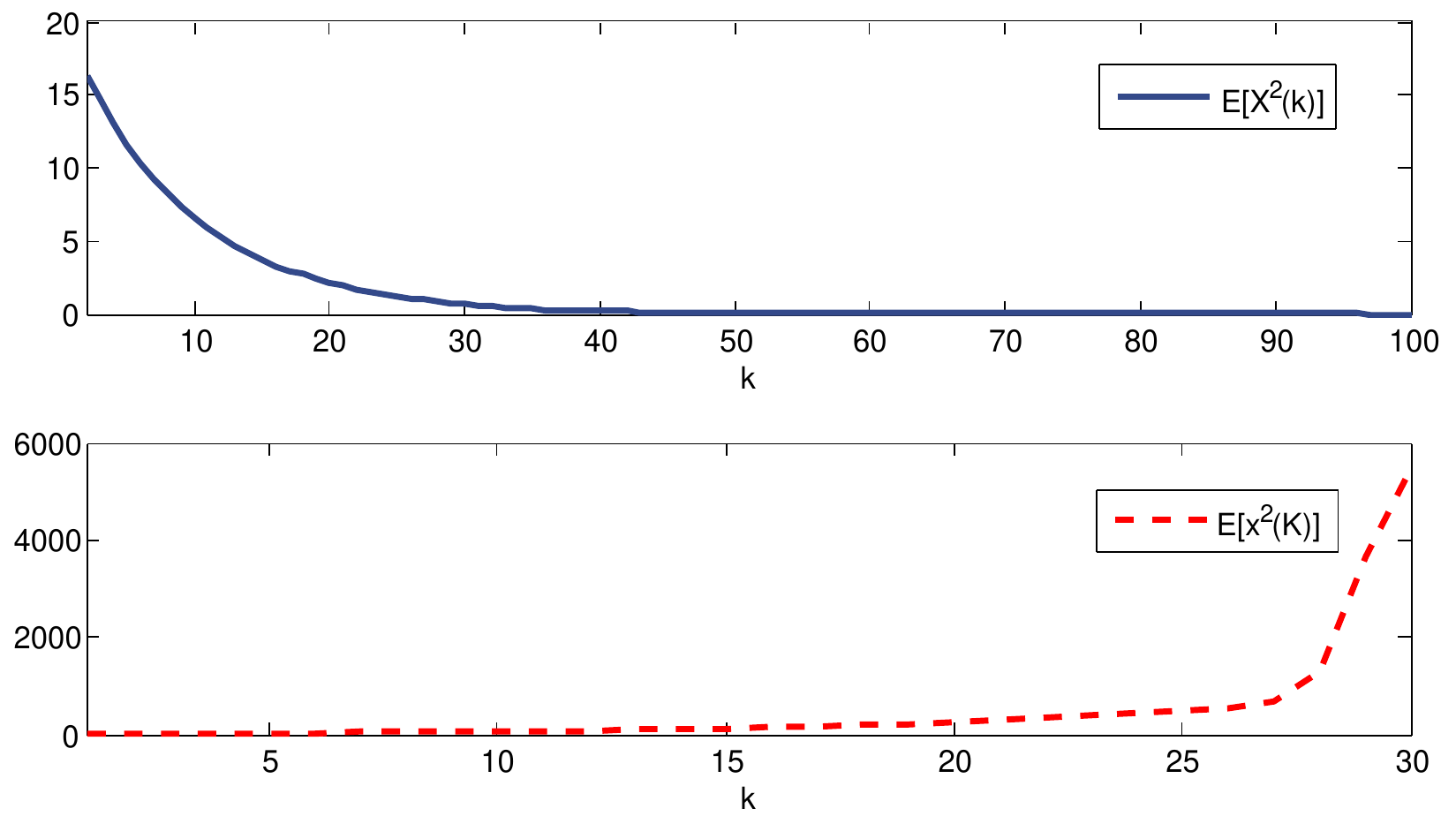}
\caption{The evolutions of $\bb{E}[\mathfrak{X}^2(k)]$ for different sample periods over an independent random network with $q=0.5$.
In the upper figure, $\mathfrak{X}^2(k)]$ converges to
$0$ as $k\to \infty$ when $\tau_\ast=1$.
In the bottom figure, $\mathfrak{X}^2(k)]$ diverges as
$k\to \infty$ when
$\tau_\ast=1.14$.}\label{fig:graph-2}
\end{center}
\end{figure}

\subsection{Mean-square Convergence vs. Divergence}
We consider a network consisting of $\sfN=4$ nodes indexed by
$\mathrm{V}=\{v_1,v_2,v_3,v_4\}$. Let
$\mathrm{E}=\big\{(v_1,v_2),(v_2,v_3),(v_3,v_2),(v_3,v_4)\big\}$.
The underlying graph $\mathrm{G}=(\mathrm{V},\mathrm{E})$ is illustrated in
Figure~\ref{fig:graph-1}. Evidently, $\mathrm{G}$ has a directed
spanning tree.
The random variables $\mathbf{1}_{\{(j,i)\in
\mathrm{E}_k\}}$, $(j,i)\in \mathrm{E}$ and $k\in\bb{N}$, are
i.i.d. Bernoulli ones with $\bb{P}\big((j,i)\in
\mathrm{E}_k\big)=0.5$.
We choose a uniform
inter-sampling interval, i.e., $\tau_k=\tau_\ast$ for all $k$.
Then Algorithm~\eqref{eqn:evolution-xt}
is given by
\begin{equation}\label{eqn:evoluction-xk-example1}
x(t_{k+1})=\big[I-\tau_\ast{L}(k)\big]x(t_{k}).
\end{equation}
According to Theorem~\ref{prop:propostion1}, we compute that
system~\eqref{eqn:evoluction-xk-example1}
achieves consensus in mean square if and only if
$\tau_\ast\leq 1.07$. We next illustrate this conclusion using simulations.
Choose $x(t_0)=[5~2~1~1]'$,
run $10^6$ Monte Carlo simulations, and
then use the average as an approximation of
$\bb{E}[\mathfrak{X}^2(k)]$.
Figure~\ref{fig:graph-2} illustrates that
$\bb{E}[\mathfrak{X}^2(k)]$ converges to
$0$ as $k$ becomes large when $\tau_\ast=1$ and diverges
as $k$ increases when $\tau_\ast=1.14$,
validating the conclusion of Theorem~\ref{prop:propostion1}.

\begin{figure}[t!]
\begin{center}
\includegraphics[width=2.45in]{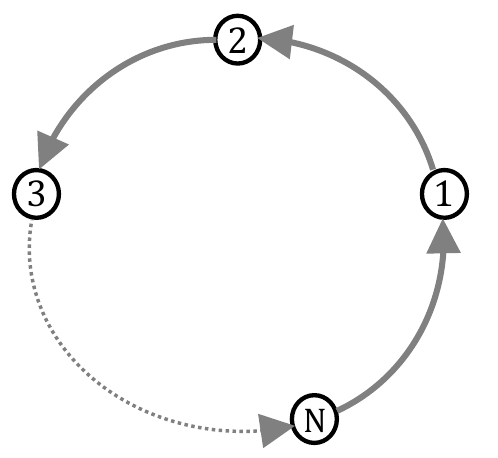}
\caption{An illustration of a directed cycle graph.}\label{fig:graph-cycle}
\end{center}
\end{figure}
\subsection{Independent and Markovian Random Graphs}
\begin{figure}[h!]
\begin{center}
\includegraphics[width=5.8in]{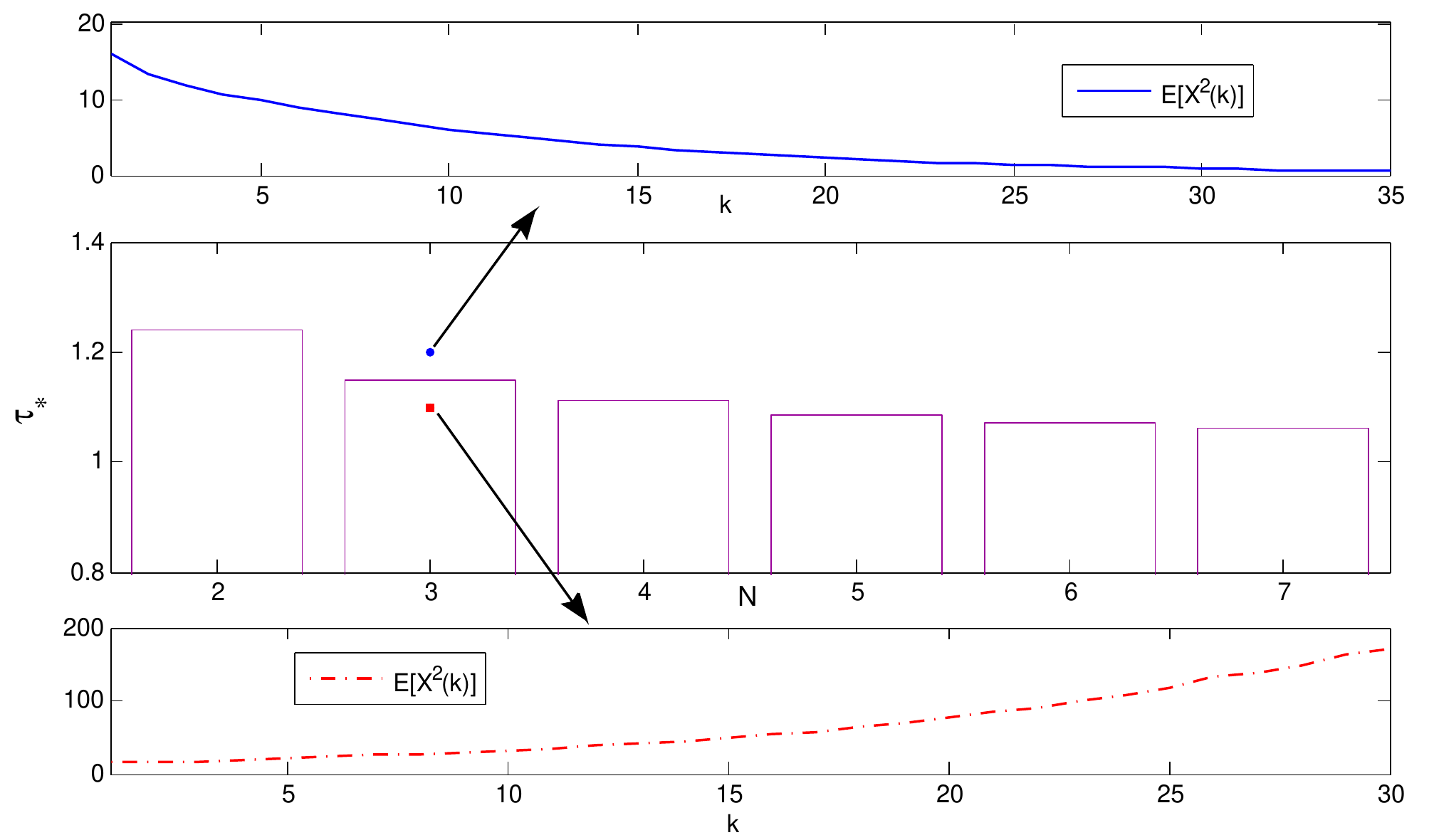}
\caption{The relationship between $\sfN$ and $\tau_\dagger$ over cycle graphs
over independent random networks ($q=0.6$).
For $\sfN=3$, two sample periods, $\tau_\ast=1.1$ (the red rectangle) and $\tau_\ast=1.2$ (the blue circle), are chosen to illustrate the divergence and convergence behaviors of $\mathbb{E}[\mathfrak{X}^2(k)]$ respectively.}\label{fig:graph-5}
\end{center}
\end{figure}

\begin{figure}[h!]
\begin{center}
\includegraphics[width=5.8in]{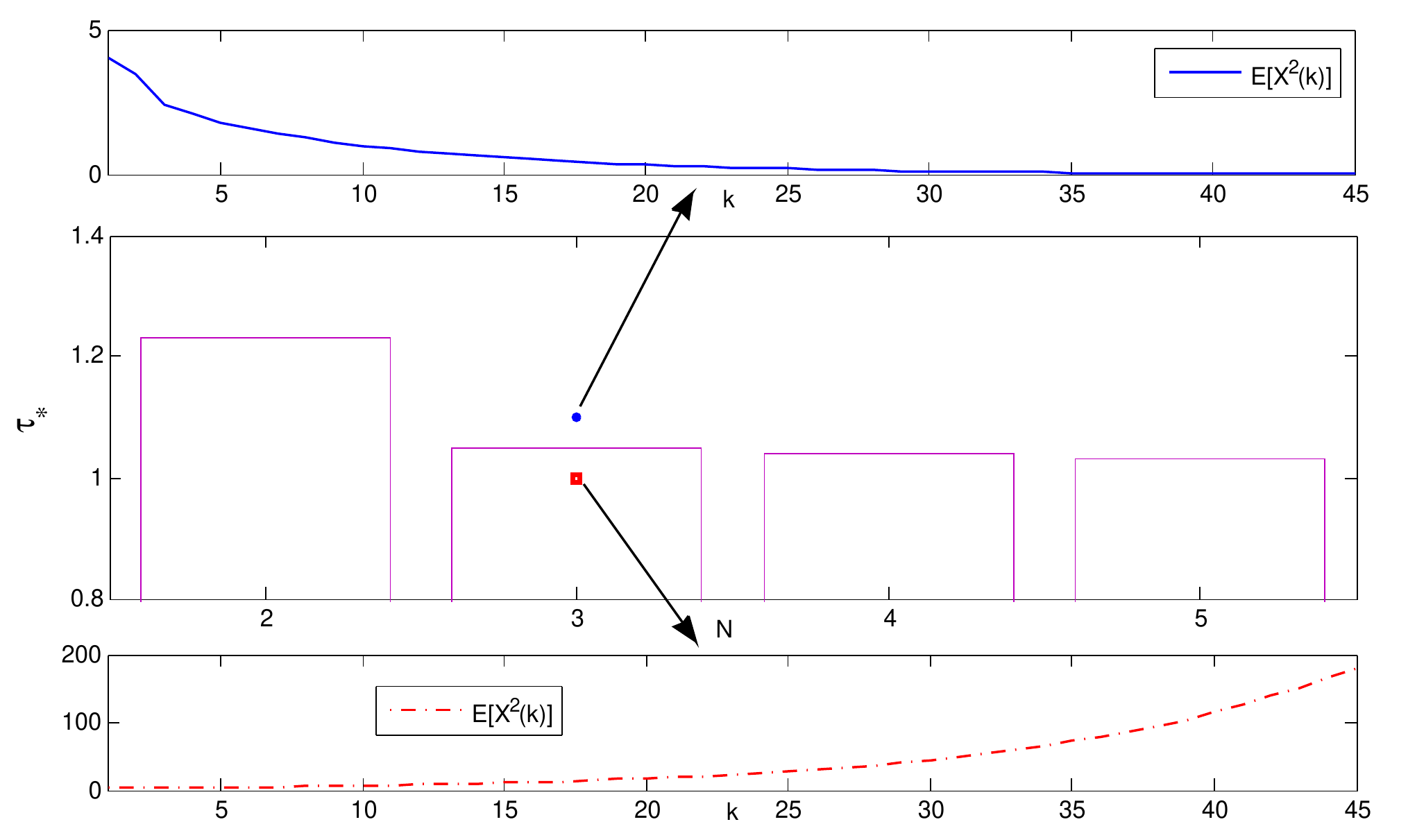}
\caption{The relationship between $\sfN$ and $\tau_\dagger$ over cycle graphs
over Markovian random networks ($p=0.6,~q=0.9$). For $\sfN=3$, two sample periods, $\tau_\ast=1.0$ (the red rectangle) and $\tau_\ast=1.1$ (the blue circle), are chosen to illustrate the divergence and convergence behaviors of $\mathbb{E}[\mathfrak{X}^2(k)]$ respectively.}\label{fig:graph-6}
\end{center}
\end{figure}

Consider a network of $\sfN$ nodes connected by a directed cycle graph as the
underlying graph, see Figure~\ref{fig:graph-cycle}. We choose $q=0.6$
for the i.i.d. model. {The relationship between the number of nodes $\sfN$ and the critical sampling interval $\tau_\dagger$ is plotted in Figure~\ref{fig:graph-5}}. As for the Markovian model,
we choose $p=0.4$ and $q=0.7$. The relationship between $\sfN$ and $\tau_\dagger$ is plotted in Figure~\ref{fig:graph-6}.
Note that each $\mathbf{1}_{\{(j,i)\in\mathrm{E}_k\}}$ has a
stationary distribution identical to the distribution of
$\mathbf{1}_{\{(j,i)\in\mathrm{E}_k\}}$ in the independent model.

\section{Conclusions}\label{sec:conclusion}
In this paper, we have considered sampled-data consensus problem over random networks.
We first defined three types of random consensus notions and established the equivalence of these consensus notions  provided a sufficient condition in terms of the inter-sampling interval and  the size of the network.
Under this condition, three types of consensus were
shown to be simultaneously achieved if the underlying graph contains a directed spanning tree.
Both independent and Markovian random networks are then
considered.
In either network model, necessary and sufficient conditions for mean-square consensus were derived in terms of the inter-sampling interval. Sufficient conditions for
almost sure convergence/divergence were also
provided, respectively, in
terms of the size of the inter-sampling interval.
The results for the independent and Markovian random networks are summarized in the following table.
\begin{table}[h]
\begin{center}
    \begin{tabular}{ |c| c| c|c |c|}
    \hline
     & \parbox[c][1.35cm][c]{2.25cm}{\centering Mean-square\\Consensue} & \parbox[c][1.35cm][c]{2.25cm}{\centering Mean-square\\Divergence}& \parbox[c][1.35cm][c]{2.25cm}{\centering Almost Sure\\Consensue} & \parbox[c][1.35cm][c]{2.25cm}{\centering Almost Sure\\Divergence} \\
    \hline
     \parbox[c][1.35cm][c]{2cm}{\centering Independent\\ Sampling} & $\tau_\ast\leq \tau_\dagger $ &  $\tau_\ast> \tau_\dagger $  & $\tau_\ast\leq \tau_\dagger $ & $\tau_\ast> \tau_\natural$\\
   \hline
     \parbox[c][1.35cm][c]{2cm}{\centering Markovian\\ Sampling} & $\tau_\ast\leq \tilde\tau_\dagger $ &  $\tau_\ast> \tilde\tau_\dagger $  & $\tau_\ast\leq \tilde\tau_\dagger $ & $\tau_\ast> \tilde\tau_\natural$\\
    \hline
    \end{tabular}
    \caption{ Summary of the consensus results for
    the independent and Markovian random networks.}
\end{center}
\end{table}
It is quite surprising that the phase transition phenomenon of mean-square
consensus exists for both types of random networks.

\bibliographystyle{IEEEtran}
\bibliography{refs,sj_reference}

\begin{thebibliography}{10}
\providecommand{\url}[1]{#1}
\csname url@samestyle\endcsname
\providecommand{\newblock}{\relax}
\providecommand{\bibinfo}[2]{#2}
\providecommand{\BIBentrySTDinterwordspacing}{\spaceskip=0pt\relax}
\providecommand{\BIBentryALTinterwordstretchfactor}{4}
\providecommand{\BIBentryALTinterwordspacing}{\spaceskip=\fontdimen2\font plus
\BIBentryALTinterwordstretchfactor\fontdimen3\font minus
  \fontdimen4\font\relax}
\providecommand{\BIBforeignlanguage}[2]{{%
\expandafter\ifx\csname l@#1\endcsname\relax
\typeout{** WARNING: IEEEtran.bst: No hyphenation pattern has been}%
\typeout{** loaded for the language `#1'. Using the pattern for}%
\typeout{** the default language instead.}%
\else
\language=\csname l@#1\endcsname
\fi
#2}}
\providecommand{\BIBdecl}{\relax}
\BIBdecl

\bibitem{JadbabaieLinMorse03}
A.~Jadbabaie, J.~Lin, and A.~S. Morse, ``Coordination of groups of mobile
  autonomous agents using nearest neighbor rules,'' \emph{{IEEE} Transactions
  on Automatic Control}, vol.~48, no.~6, pp. 988--1001, 2003.

\bibitem{SaberMurray04}
R.~Olfati-Saber and R.~M. Murray, ``Consensus problems in networks of agents
  with switching topology and time-delays,'' \emph{{IEEE} Transactions on
  Automatic Control}, vol.~49, no.~9, pp. 1520--1533, 2004.

\bibitem{LinZhiyun_SIAM07}
Z.~Lin, B.~Francis, and M.~Maggiore, ``State agreement for continuous-time
  coupled nonlinear systems,'' \emph{SIAM Journal of Control and Optimization},
  vol.~46, no.~1, pp. 288--307, 2007.

\bibitem{Boyd_SCL04}
L.~Xiao and S.~Boyd, ``Fast linear iterations for distributed averaging,''
  \emph{Systems and Control Letters}, vol.~53, pp. 65--78, 2004.

\bibitem{RenBeard05_TAC}
W.~Ren and R.~W. Beard, ``Consensus seeking in multiagent systems under
  dynamically changing interaction topologies,'' \emph{{IEEE} Transactions on
  Automatic Control}, vol.~50, no.~5, pp. 655--661, 2005.

\bibitem{CaoYongcan_IJRNC10}
Y.~Cao and W.~Ren, ``Multi-vehicle coordination for double-integrator dynamics
  under fixed undirected/directed interaction in a sampled-data setting,''
  \emph{International Journal of Robust and Nonlinear Control}, vol.~20, pp.
  987--1000, 2010.

\bibitem{gao2011sampled}
Y.~Gao and L.~Wang, ``Sampled-data based consensus of continuous-time
  multi-agent systems with time-varying topology,'' \emph{IEEE Transactions on
  Automatic Control}, vol.~56, no.~5, pp. 1226--1231, 2011.

\bibitem{zhang2010consensus}
Y.~Zhang and Y.-P. Tian, ``Consensus of data-sampled multi-agent systems with
  random communication delay and packet loss,'' \emph{IEEE Transactions on
  Automatic Control}, vol.~55, no.~4, pp. 939--943, 2010.

\bibitem{Feng-TAC12}
F.~Xiao and T.~Chen, ``Sampled-data consensus for multiple double integrators
  with arbitrary sampling,'' \emph{IEEE Transactions on Automatic Control},
  vol.~57, no.~12, pp. 3230--3235, 2012.

\bibitem{Kar-TSP08}
S.~Kar and J.~M. Moura, ``Sensor networks with random links: topology design
  for distributed consensus,'' \emph{IEEE Transactions on Signal Processing},
  vol.~56, no.~7, pp. 3315--3326, 2008.

\bibitem{Kar-TSP09}
------, ``Distributed consensus algorithms in sensor networks: quantized data
  and random link failures,'' \emph{IEEE Transactions on Signal Processing},
  vol.~58, no.~3, pp. 1383--1400, 2009.

\bibitem{pereira-TSP2010}
S.~S. Pereira and A.~Pag\'es-Zamora, ``Mean square convergence of consensus
  algorithms in random {WSN}s,'' \emph{IEEE Transactions on Signal Processing},
  vol.~58, no.~5, pp. 2866--2874, 2010.

\bibitem{fagnani2008randomized}
F.~Fagnani and S.~Zampieri, ``Randomized consensus algorithms over large scale
  networks,'' \emph{IEEE Journal on Selected Areas in Communications}, vol.~26,
  no.~4, pp. 634--649, 2008.

\bibitem{Fagnani-SIAM}
------, ``Average consensus with packet drop communication,'' \emph{SIAM
  Journal on Control and Optimization}, vol.~48, no.~1, pp. 102--133, 2009.

\bibitem{ali-TAC10}
A.~Tahbaz-Salehi and A.~Jadbabaie, ``Consensus over ergodic stationary graph
  processes,'' \emph{IEEE Transactions on Automatic Control}, vol.~55, no.~1,
  pp. 225--230, 2010.

\bibitem{Baras-SIAM13}
I.~Matei, J.~S. Baras, and C.~Somarakis, ``Convergence results for the linear
  consensus problem under markovian random graphs,'' \emph{SIAM Journal on
  Control and Optimization}, vol.~51, no.~2, pp. 1574--1591, 2013.

\bibitem{chen-TAC2011}
Q.~Song, G.~Chen, and D.~W. Ho, ``On the equivalence and condition of different
  consensus over a random network generated by i.i.d. stochastic matrices,''
  \emph{IEEE Transactions on Automatic Control}, vol.~56, no.~5, pp.
  1203--1207, 2011.

\bibitem{mesbahi-TAC05}
Y.~Hatano and M.~Mesbahi, ``Agreement over random networks,'' \emph{{IEEE}
  Transactions on Automatic Control}, vol.~50, no.~11, pp. 1876--1872, 2005.

\bibitem{Wu_TAC06}
C.~W. Wu, ``Synchronization and convergence of linear dynamics in random
  directed networks,'' \emph{IEEE Transactions on Automatic Control}, vol.~51,
  no.~7, pp. 1207--1210, 2006.

\bibitem{Stilwell-TAC07}
M.~Porfiri and D.~Stilwell, ``Consensus seeking over random weighted directed
  graphs,'' \emph{IEEE Transactions on Automatic Control}, vol.~51, no.~7, pp.
  1767--1773, 2007.

\bibitem{ali-TAC08}
A.~Tahbaz-Salehi and A.~Jadbabaie, ``A necessary and sufficient condition for
  consensus over random networks,'' \emph{IEEE Transactions on Automatic
  Control}, vol.~53, no.~3, pp. 791--795, 2008.

\bibitem{acemoglu2013opinion}
D.~Acemoglu, G.~Como, F.~Fagnani, and A.~Ozdaglar, ``Opinion fluctuations and
  disagreement in social networks,'' \emph{Mathematics of Operations Research},
  vol.~38, no.~1, pp. 1--27, 2013.

\bibitem{shi2013agreement}
G.~Shi, M.~Johansson, and K.~H. Johansson, ``How agreement and disagreement
  evolve over random dynamic networks,'' \emph{IEEE Journal on Selected Areas
  in Communications}, vol.~31, no.~6, pp. 1061--1071, 2013.

\bibitem{FengLong_TAC08}
F.~Xiao and L.~Wang, ``Asynchronous consensus in continuous-time multi-agent
  systems with switching topology and time-varying delays,'' \emph{IEEE
  Transactions on Automatic Control}, vol.~53, no.~8, pp. 1804--1816, 2008.

\bibitem{hardy1952inequalities}
G.~H. Hardy, J.~E. Littlewood, and G.~P{\'o}lya, \emph{Inequalities}.\hskip 1em
  plus 0.5em minus 0.4em\relax Cambridge {U}niversity {P}ress, 1952.

\bibitem{durrett2010probability}
R.~Durrett, \emph{Probability: Theory and Examples}.\hskip 1em plus 0.5em minus
  0.4em\relax Cambridge {U}niversity {P}ress, 2010.

\bibitem{Guanrong11tac}
Q.~Song, G.~Chen, and D.~Ho, ``On the equivalence and condition of different
  consensus over a random network generated by i.i.d. stochastic matrices,''
  \emph{IEEE Transactions on Automatic Control}, vol.~56, no.~5, pp.
  1203--1207, 2011.

\bibitem{Costa04TAC}
O.~L.~V. Costa and M.~Fragoso, ``Comments on ''stochastic stability of jump
  linear systems'','' \emph{IEEE Transactions on Automatic Control}, vol.~49,
  no.~8, pp. 1414--1416, 2004.

\bibitem{boyd1994lmi}
S.~P. Boyd, L.~El~Ghaoui, E.~Feron, and V.~Balakrishnan, \emph{Linear matrix
  inequalities in system and control theory}.\hskip 1em plus 0.5em minus
  0.4em\relax SIAM, 1994.

\bibitem{horn2012matrix}
R.~A. Horn and C.~R. Johnson, \emph{Matrix analysis}.\hskip 1em plus 0.5em
  minus 0.4em\relax Cambridge {U}niversity {P}ress, 2012.

\bibitem{feller1968introduction}
W.~Feller, \emph{An Introduction to Probability Theory and its Applications
  Vol. {I}}.\hskip 1em plus 0.5em minus 0.4em\relax John Wiley \& Sons, 1950.

\bibitem{rudin1987real}
W.~Rudin, \emph{Real and Complex Analysis}.\hskip 1em plus 0.5em minus
  0.4em\relax Tata McGraw-Hill Education, 1987.

\bibitem{costa2006discrete}
O.~L.~V. Costa, M.~D. Fragoso, and R.~P. Marques, \emph{Discrete-time {M}arkov
  jump linear systems}.\hskip 1em plus 0.5em minus 0.4em\relax Springer Science
  \& Business Media, 2006.

\end{thebibliography}

\medskip

\medskip

\medskip

\noindent {\sc Junfeng Wu, Tao Yang, and Karl H. Johansson} \\
{\noindent {\small  ACCESS Linnaeus Centre,
   School of Electrical Engineering,
KTH Royal Institute of Technology,
\\ Stockholm 100 44, Sweden }\\}
       {\small Email: {\tt\small junfengw@kth.se, taoyang@kth.se, kallej@kth.se}

\medskip

\noindent {\sc Ziyang Meng} \\
{\noindent {\small   Institute for Information-Oriented Control, Technische Universitat Munchen, \\
D-80290 Munich, Germany }\\}
       {\small Email: {\tt\small ziyang.meng@tum.de}

\medskip


\medskip

{\noindent {\sc Guodong Shi}} \\
{\noindent {\small College of Engineering and Computer Science, The Australian National University, \\ Canberra, ACT 0200 Australia}}\\  {\small Email: } {\tt\small guodong.shi@anu.edu.au}

\end{document}